\documentclass[graphicx,preprint,11pt]{aastex}

\usepackage{color}

\lefthead{}
\righthead{}

\begin{document}

\title{The timeline of the Lunar bombardment - revisited.} 

\author{{A. Morbidelli$^{(1)}$, D.Nesvorny$^{(2)}$, V. Laurenz$^{(3)}$, S. Marchi$^{(2)}$, D.C. Rubie$^{(3)}$, L. Elkins-Tanton$^{(4)}$ M. Wieczorek$^{(1)}$ and S.Jacobson $^{(1,3,5)}$}       \\  
(1) Laboratoire Lagrange, UMR7293, Universit\'e C\^ote d'Azur,
  CNRS, Observatoire de la C\^ote d'Azur. Boulevard de l'Observatoire,
  06304 Nice Cedex 4, France. (Email: morby@oca.eu / Fax:
  +33-4-92003118) \\
(2) Soutwest Research Institute, Boulder, Co.  \\ 
(3) Bayerisches Geoinstitut, Bayreuth, Germany\\
(4) School of Earth and Space exploration, University of Arizona, Phoenix, Az.\\
(5) Department Of Earth And Planetary Sciences, Northwestern University} 

\begin{abstract}
The timeline of the lunar bombardment in the first Gy of Solar System history remains  {unclear. Basin-forming impacts (e.g. Imbrium, Orientale), occurred 3.9--3.7~Gy ago, i.e. 600-800~My after the formation of the Moon itself. Many other basins formed before Imbrium, but their exact ages are not precisely known. There is an intense debate between two possible interpretations of the data: in the cataclysm scenario there was a surge in the impact rate approximately at the time of Imbrium formation, while in the accretion tail scenario the lunar bombardment declined since the era of planet formation and the latest basins formed in its tail-end.} Here, we revisit the work of Morbidelli et al. (2012) that examined which scenario could be compatible with both the lunar crater record in the 3--4~Gy period and the abundance of highly siderophile elements (HSE) in the lunar mantle. We use updated numerical simulations of the fluxes of asteroids, comets and planetesimals leftover from the planet-formation process. Under the traditional assumption that the HSEs track the total amount of material accreted by the Moon since its formation, we conclude that only the cataclysm scenario can explain the data. The cataclysm should have started $\sim 3.95$~Gy ago. However we also consider the possibility that HSEs are sequestered from the mantle of a planet during magma ocean crystallization, due to iron sulfide exsolution (O'Neil, 1991; Rubie et al., 2016).  We show that this is likely true also for the Moon, {if mantle overturn is taken into account}. Based on the hypothesis that the lunar magma ocean crystallized about 100-150~My after Moon formation (Elkins-Tanton et al., 2011), and therefore that HSEs accumulated in the lunar mantle only after this timespan, we show that { the bombardment in the 3--4~Gy period} can be explained in the accretion tail scenario. This hypothesis would also explain why the Moon appears so depleted in HSEs relative to the Earth. We also extend our analysis of the cataclysm and accretion tail scenarios to the case of Mars. The accretion tail scenario requires a global resurfacing event on Mars $\sim 4.4$Gy ago, possibly associated with the formation of the Borealis basin, and it is consistent with the HSE budget of the planet. Moreover it implies that the Noachian and pre-Noachian terrains are $\sim 200$~My older than usually considered. 
\end{abstract}

\section{Introduction}

Soon after the Apollo missions returned lunar rocks, it was realized (Papanastassiou and Wasserburg, 1971a, 1971b; Wasserburg and Papanastassiou, 1971; Turner
et al., 1973) that many of them carry evidence for impact shocks that occurred about 3.9 Gy ago. Thus the Moon experienced a heavy bombardment about 600 My after planet formation during which the youngest lunar basins, like Imbrium and Orientale (and possibly also Serenitatis) formed (for reviews see e.g. Hartmann et al., 2000;  Chapman et al., 2007; Norman, 2009). We call this period of intense bombardment (more intense than the one in the current eon of the Solar System) the {\it Late Heavy Bombardment} (LHB). The existence of a LHB is not in dispute; the name itself does not carry any prejudice on what was its cause. 

Instead, two contrasting views exist on the origin of the LHB and on the overall timeline of the lunar bombardment in general. 

One view is that of the {\it Lunar terminal cataclysm}. This name was introduced in Tera et al. (1974) who, in order to explain why signatures for impacts older than 3.9~Gy were virtually absent, suggested that the LHB was the consequence of a prominent spike in the impact rate. Interestingly, the impact age distributions of some meteorites from the asteroid belt are somewhat similar to that of the Moon, indicating that the cataclysm may have affected the whole inner solar system (Marchi et al., 2013). Among the prominent works that advocated the cataclysmic scenario are Ryder (1990, 2002),  Cohen et al. (2000), St{\"o}ffler and Ryder (2001) and  Marchi et al. (2012). {Some authors (including some co-authors of this paper in the past) used the term LHB for the cataclysm. We are cautious to avoid this confusion here.}

The opposite view is that of the {\it accretion tail}. In this view, the lunar bombardment decayed monotonically since the time of formation of the terrestrial planets (roughly 4.5 Gy ago), and the apparent concentration of impact ages around 3.9~Gy just reflects sampling biases or burial of the oldest rocks (Hartmann, 1975, 2003; {Haskin et al. 1998, 2003}) or age resetting (Boehnke and Harrison, 2016). The recent discovery of a basin-forming event 4.2~Gy ago (Norman and Nemchin, 2014) seems to support this view (but not necessarily inconsistent with the most modern views of the cataclysm scenario: Morbidelli et al., 2012; Marchi et al., 2013). Among other prominent works that advocated the accretion tail scenario are Neukum et al. (2001), Ivanov (2001) and Werner (2014). 

Unfortunately, the lunar crater record does not allow discriminating unambiguously these two views of the lunar bombardment history. This is because the surface units with well determined radiometric ages are just a few and they are all younger than 3.9 Gy (Neukum and Wilhelms, 1982; Marchi et al., 2009; Robbins, 2014). In the past, it was believed that the Nectaris basin unit could be used as a reference for the bombardment 4.1~Gy ago (Maurer et al., 1978; Neukum and Ivanov, 1994), but later work (Norman et al., 2010) showed that the actual age of the Nectaris basin is highly uncertain. 

The accretion tail hypothesis fits the straightforward expectations on the evolution of the bombardment of planets that formed from a planetesimal disk: the planetesimals leftover from the main planet-formation period are progressively removed by a combination of collisions and dynamical effects, so that the bombardment that they cause wanes over time.  Instead, for a long time the main problem for the cataclysm hypothesis was the lack of a plausible explanation for the impact surge. However, in 2005 it was proposed that the giant planets of the Solar System underwent a dynamical instability (Tsiganis et al., 2005; Morbidelli et al., 2005). It was shown that, under specific conditions, such an instability could have occurred hundreds of My after the main era of planet formation (Gomes et al., 2005). The dispersal of planetesimal populations caused by a late planet instability could have produced the putative cataclysm. 

While the evidence for a giant planet instability strengthened over the years with the detailed analysis of the Solar System structure (Nesvorny et al., 2007; Nesvorny and Morbidelli, 2012; Brasser and Morbidelli, 2013; Nesvorny et al. 2013; Nesvorny, 2015a,b), the dynamical models remained agnostic as to whether the instability occurred early or late. Both solutions remain possible depending on the properties of the original trans-Neptunian disk (Levison et al., 2011; Deienno et al., 2017). 

Morbidelli et al. (2012; M12 hereafter), attempted to constrain the timing of the giant planet instability. Following Ryder (2002), they tried to reconcile the LHB with the total mass accreted by the Moon since its formation, recorded in the abundance of the highly siderophile elements (HSE) in {rocks derived from the lunar mantle} (Day et al., 2007; Walker, 2009, 2014; Day and Walker, 2015). The argument rested upon the traditional assumption that HSEs are removed from the mantle only by metal segregation during core formation. Because core formation and lunar formation are expected to be coeval, the HSEs would track the amount of chondritic material that hit the Moon since its formation. The lunar mantle is extremely depleted in HSEs, implying that the total mass accreted by the Moon was only $\sim 2.5\times 10^{-6}$ Earth Masses ($M_\oplus$) (Day et al., 2007; Day and Walker, 2015). The total amount of HSEs in the lunar crust is just $\sim 1/4$ of that in the lunar mantle content (Ryder, 2002) and can be neglected for mass balance purposes, given the uncertainties.  {This tight constraint on the total accreted mass, translates into an upper bound on the leftover planetesimal population. M12 showed that this upper bound is inconsistent with the cratering rate observed at the time of the LHB.} So M12 concluded that the only possible explanation for the LHB was the late injection of fresh projectiles due to a late planetary instability, i.e. a cataclysm. Using simulations of the bombardment of the Moon by asteroids destabilized onto planet-crossing orbits  by the giant planet instability from an extended portion of the inner asteroid belt (Bottke et al., 2012), M12 concluded that the giant planet instability should have occurred about 4.1~Gy ago. The LHB would have accounted for about 1/4 to 1/3 of the total number of basins, or about 12 basins. The others would have been formed in an accretion tail, mostly before 4.2~Gy ago. 

There are two reasons to revisit in depth the work of M12. First, a new result (Rubie et al., 2016), based on an original idea by O'Neill (1991), shows that {there is widespread HSE sequestration into the core of a planetary body during magma ocean crystallization, due to the exsolution and segregation of liquid FeS from the crystallizing silicate}. This implies that the current HSE concentrations in the lunar mantle may only record the amount of chondritic material that was delivered to the Moon after the crystallization of its magma ocean. The difference is significant, because the lunar magma ocean is estimated to have crystallized up to $\sim$200 My after Moon formation, due to strong tidal heating and the development of an insulating feldspathic crust (Elkins-Tanton et al., 2011). Thus, it is possible that the total mass impacting the Moon after its formation was significantly larger than that considered in M12, thereby changing profoundly their conclusions. 

The second reason is more technical but nevertheless important: new simulations are now available on the flux of asteroids from the main belt to the terrestrial-planet region, before and after the giant planet instability (Nesvorny et al., 2017). These simulations are superior to those of Bottke et al. (2012), used in M12, because they enact a much more constrained and realistic scenario of giant planet instability. Moreover, they follow all asteroids, not just those of the extended inner-belt, and for the full lifetime of the Solar System. This point is important because it allows  a more reliable calibration of the original population based on the surviving one, namely the current asteroid population. Nesvorny et al. (2017) announced that the number of basin-forming projectiles from the asteroid belt is much smaller than that reported in Bottke et al. (2012), so the use of these new simulations may also change substantially the conclusions of M12.

With the goal of revisiting the M12 work in depth, this paper is structured as follows. In section~\ref{NewSims} we briefly describe the Nesvorny et al. (2017) simulations and how we link them to the formation of craters of a given size as a function of time. We also include the role of comets in the lunar bombardment, destabilized from the trans-Neptunian disk, and leftover planetesimals from the terrestrial planet formation era. {This technical section can be skipped without loosing the global picture}. In section~\ref{Revisit} we redo the M12 calculations, but using the new simulations. That is, we assume again that the total mass accreted by the Moon since its formation is that recorded by the lunar mantle HSEs, and we attempt to reconcile this mass with the LHB record. We conclude again for the need of a cataclysm, but with properties quite different from those deduced in the original work. In section~\ref{Lunar}, we discuss the exsolution of FeS during the inside-out crystallization of the lunar magma ocean and its subsequent transport to the core/mantle boundary during mantle overturn. We conclude that the HSEs currently present in the lunar mantle are those delivered by chondritic material {\it after} mantle crystallization and overturn. In section~\ref{New}, we use this conclusion to show that, if the mantle overturn happened $\sim 4.4$~Gy ago, the LHB can be explained in the accretion tail scenario. This is the first time that the accretion tail scenario is shown to be potentially consistent with both the lunar HSE concentrations and the LHB crater record. In section~\ref{Earth-Moon}, we discuss another advantage of the idea that the accumulation of HSEs in the upper lunar mantle started late. In fact, it can explain why the Moon appears to be much more depleted in HSEs than the Earth (Bottke et al., 2010). In section~\ref{Mars} we extend our analysis to the bombardment of Mars, in order to gain additional insight that can help to discriminate between the cataclysm and accretion tail scenarios. Finally, in section~\ref{conclusions} we summarize our results and discuss the advantages and disadvantages of each scenario.

\section{New numerical simulations of the evolution of small bodies in the aftermath of terrestrial planet formation and giant planet instability}
\label{NewSims}

In order to assess the asteroidal bombardment triggered by the instability of the giant planets we use the results of the simulations labeled CASE1B in the paper of Nesvorny et al. (2017).  The number of asteroidal collisions with the Earth-Moon system since the beginning of the instability declines with time as shown by Fig.~11b of that paper (see also figures~\ref{miniCat} and~\ref{noCat} in this paper, discussed later). The total collision probability with the Moon per asteroid initially in the main belt is $5.5\times 10^{-4}$. The ratio of collision probabilities with Mars and the Moon is $P^{ast}_{Mars/Moon} =7.3$. In a late instability scenario, the simulations of Nesvorny et al. (2017) show that the number of impacts over a time period of 400~My prior to the instability is roughly equal to that recorded in the simulation starting with the giant planet instability (see their Fig.~11b). In the simulations of Bottke et al. (2012), however, the number of collisions recorded in the pre-instability phase was only 10-20\%. The reason for this difference is unclear. Thus, taking the average between these two results, we assumed in this work that the total number of impacts in the pre-instability phase is 60\% of that occurring after the instability, and distributed uniformly in time. These assumptions are not crucial for the conclusions of this paper.  

For the comets coming from the trans-Neptunian disk, we used the output from the simulations in Nesvorny et al. (2013) and we computed the collision probabilities with the terrestrial planets (not included in the simulation but assumed to be on their current orbits) using an Opik-like algorithm, recently improved in Vokrouhlicky et al. (2012) to treat accurately also the case of intersecting orbits with a large mutual inclination.  The collision probability with the Earth-Moon system declines with time much more steeply than the asteroid curve, because comets are rapidly ejected by close encounters with Jupiter (Fig.~\ref{miniCat}). Over the dynamical lifetime, the total collision probability with the Moon per comet initially in the trans-Neptunian disk is $1.8\times 10^{-8}$. The ratio of collision probabilities with Mars and the Moon is $P^{com}_{Mars/Moon} =3.5$.

\begin{figure}[t!]
\centerline{\includegraphics[width=9.cm]{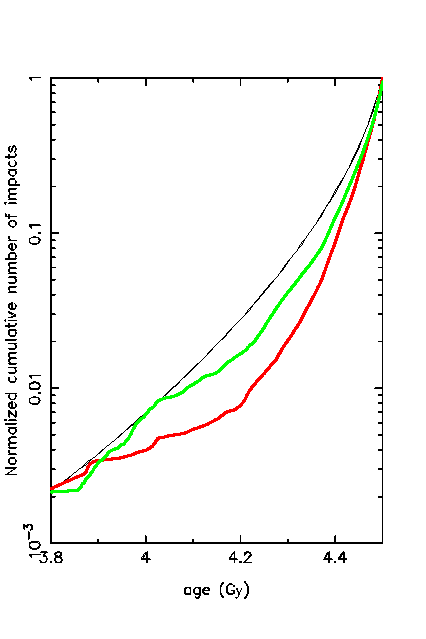}}
\caption{\footnotesize The cumulative number of impacts generated on the Earth-Moon system as a function of age (here normalized to the total number of impacts since a hypothetical planet-formation time of 4.5~Gy, but notice that the Moon could have formed up to $\sim 4.45$~Gy ago; Jacobson et al., 2014). The red  and green  thick curves show two simulations of the same process. The black thin curve shows the integral of the function $C_1(t)$ defined in the text. In the simulations, the initial planetesimal distribution is taken from the simulation presented in Walsh et al. (2011) at two different times: 30 and 50~My (the terrestrial planets in the considered Walsh et al. simulation formed within 30~My). The evolution of the planetesimals is simulated assuming that all planets, terrestrial and giants are on their current orbital configuration. The number of impacts have been computed using the Vokrouhlicky et al. (2012) code. }
\label{newDecay}
\end{figure}

For the decay of the population of leftover planetesimals, M12 conducted eight simulations in the framework of the terrestrial planet formation model of Walsh et al. (2011). These simulations gave somewhat different decay rates of the number of collisions produced on the Earth-Moon system. Over the first 0.5~Gy, the observed numbers of collisions per unit time were bounded by the curves  $C_1(t)=\exp[-(4500-t)/10]^{0.5}$ and $C_2(t)=\exp[-(4500-t)/3]^{0.34}$ where $t$ is in My. Brasser et al. (2016) performed similar simulations and fitted the decay of the number of collisions per unit time by the function $C_3(t)=\exp[-(4500-t)/12]^{0.44}$, which is somewhat shallower. 

All these simulations had been conducted assuming that the giant planets were at the time on more circular and coplanar orbits, which is appropriate if these planets underwent a dynamical instability and acquired their current orbits at a later time.  However, because in section~\ref{New} we are going to argue in favor of an early instability of the giant planet system, we have performed two new simulations of the decay of the leftover planetesimal population assuming that all planets, giant and terrestrial, were already on the present configuration. The decay of the impact rate on the Earth-Moon system is shown in cumulative form in Fig~\ref{newDecay}. In both simulations, the decay of the number of collisions is initially steeper than predicted by the function $C_1(t)$ (the steepest among those reported above), but then it shallows off. This steeper evolution is expected, because the giant planets are more eccentric and therefore their resonances are more effective in removing bodies. However, after 500-700~My, the number of collisions agrees very well with the analytic function. Because 500-700~My after the beginning of the simulation corresponds to the time period ranging from 4.0 to 3.8~Gy ago, namely the LHB period, we will adopt $C_1(t)$ in the rest of this paper for simplicity. In fact, the actual bombardment rate in between 4.5~Gy and 4.0~Gy is unconstrained by data. Thus, what is important in the following is just the ratio between the total number of impacts since Moon formation and the number of impacts during the LHB (i.e. younger than 4~Gy), which is the same for the $C_1$ function and for the red and blue curves shown in Fig.~\ref{newDecay}.  

In the original simulations of M12, as well as in those presented in Fig.~\ref{newDecay}, the ratio of collision probabilities with Mars and the Moon is $P^{pl}_{Mars/Moon} =2$, i.e. Mars suffers 1/2 of the impacts of the Moon per unit surface given that its surface is 4 times larger. 

\subsection{Calibration of the asteroid and comet populations}

In order to compute the number of craters formed by asteroids and comets on a terrestrial body, we need an estimate of the total number of objects in these populations.

In Nesvorny et al. (2017) the fraction of the initial population surviving in the asteroid belt was tracked until the end of the 4.5~Gy simulation. This population was imposed to be equal to the number of asteroids existing today larger than a reference size of 10~km, not belonging to any known asteroid family. With this calibration, Nesvorny et al. found that the total number of asteroids with $D>10$~km impacting the Moon after the giant planet instability is $\sim 8$.

For the comets, we use the fact that in the simulations of Nesvorny et al. (2013) the fraction of the trans-Neptunian disk objects captured in the Trojan region is $(6-8) \times 10^{-7}$. This implies that the ratio between the probabilities of capture as Trojan or collision with the Moon is 39. In other words, the largest comet impacting the Moon should have a size comparable to the 39th largest Trojan. The 39th brightest Trojan has absolute magnitude 9.2. The main uncertainty comes from the albedo, which we assumed to be $0.065\pm 0.025$ (Fernandez et al., 2009), giving a preferred size of 75~km, with an uncertainty ranging from 64 to 93~km. 

\subsection{Scaling laws}

In this work we will consider two reference sizes for craters on the Moon: $D=1$~km and $D=20$~km and a reference size for large Martian craters $D=150$~km. It is essential to know which projectile are needed to produce craters of these sizes.

\begin{figure}[t!]
\centerline{\includegraphics[width=9.cm]{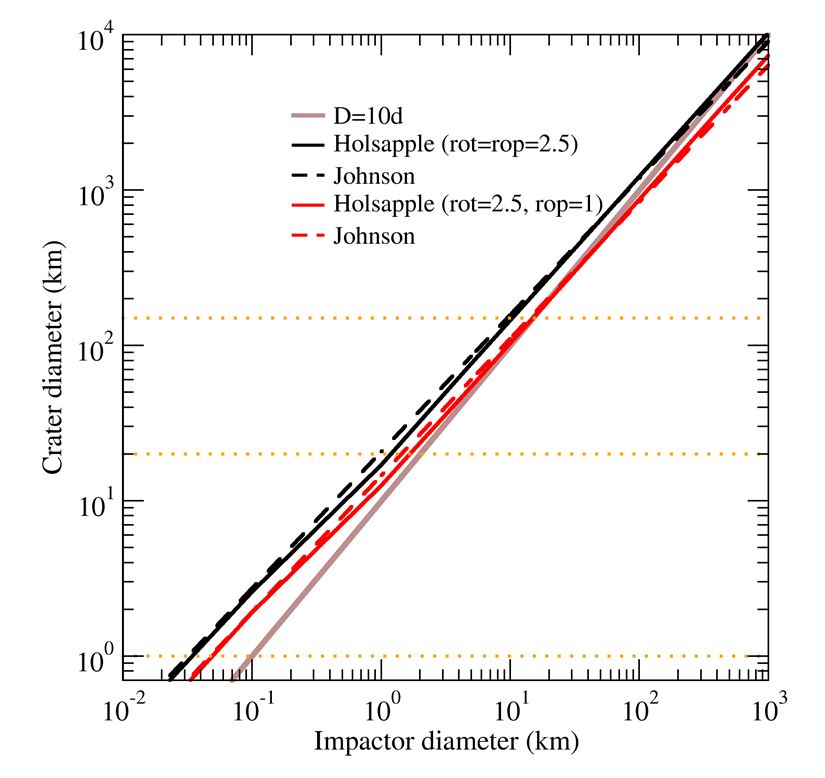}\,\includegraphics[width=9.cm]{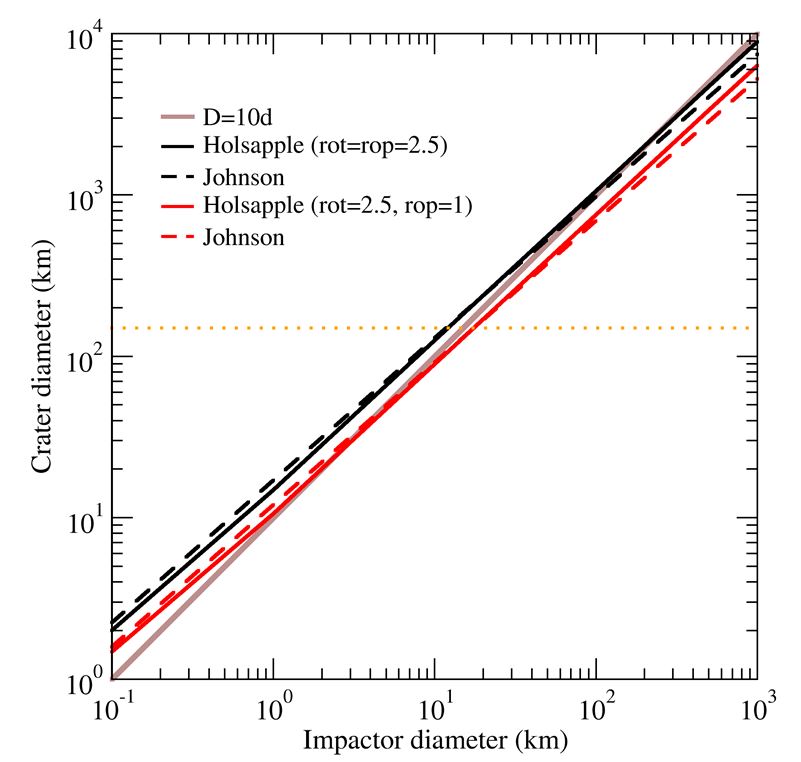} }
\caption{\footnotesize Scaling laws (Holsapple and Housen, 2007; Johnson et al., 2016) for the Moon (left panel) and Mars (right panel), for a target density $\rho_t=2.5$~g/cm$^{3}$ and projectile density $\rho_p=2.5$ and 1~g/cm$^{3}$. The assumed impact velocities are of 18~km/s for the Moon and 14~km/s for Mars, with an impact angle of 45 degrees. For reference, the relationship $D_{crater}=10 D_{impactor}$ is also plotted. }
\label{scalings}
\end{figure}

We use the scaling law between projectile size and crater size given in (Holsapple and Housen, 2007), which is very similar to that in Johnson et al. (2016). These scaling laws are reported in Fig.~\ref{scalings}.
 
We assume collision velocities of 18~km/s on the Moon and 14~km/s on Mars, consistent with numerical simulations. We assume {the most frequent} impact angle of 45 degrees and a projectile density of 2.5 g/cm$^{3}$. With these inputs, we find that a $D=1$~km crater on the Moon would be formed by a $D=50$~m object, a $D=20$~km crater by a $D=1$~km object and a $D=150$~km basin on Mars by a $D=12$~km object. The numbers above apply to rocky impactors. Cometary impactors, however, may have a significantly reduced density. The effect of the impactor's density is also shown in Fig.~\ref{scalings}. For instance, all other parameters being the same, a $D=150$~km on Mars requires an impactor with density 1 g/cm$^{3}$ to have a diameter of 18~km. 
 
\subsection{Size distributions}

We assume that all projectile populations have size distributions analogous to that of the main asteroid belt. This is supported by the observation that the size distribution of craters on the Moon is very similar to that expected to be produced by the size distribution of main belt asteroids (Strom et al., 2005), even if not identical (Minton et al., 2015). As for comets, the results of the New Horizon mission show that trans-Neptunian objects smaller than 100~km in diameter have also an asteroid-like size distribution, at least down to 1--2~km (Singer et al., 2016). 

Table~I reports the number of main belt asteroids larger than some reference sizes that we use in this paper, taken from Bottke et al. (2005).

\begin{table}[t!]
\caption{Number of main belt asteroids larger than a reference size. From Bottke et al. (2005)}
\begin{tabular}{|c|c|}
\\\hline\\
 Reference diameter $D$ (km)  & Number of asteroids larger than $D$; $n(>D)$\cr

\\\hline\\
1 &  1,100,000 \cr
10 & 8,000 \cr
12 & 5,400 \cr
18 & 2,500 \cr
75 &  370 \cr
900 & 1\cr
 \\\hline \\
\end{tabular}
\label{sjs}
\end{table}

Below 1~km the size distribution of the asteroid belt is not well constrained. However, we have an empirical measure on the Moon of the density ratio of craters larger than $D=1$~km (Neukum and Ivanov, 1984) and $D=20$~km (Marchi et al., 2012): $N_{1/20}=1400$ .

With these numbers, we have all the required elements to obtain the numbers reported in this paper. For instance, knowing that 8 asteroids larger than 10~km impact the Moon since the giant planet instability, the number of $D>1$~km craters produced will be:
$$
N_1=8 n(>1)/n(>10) N_{1/20}=1.5\times 10^6
$$
which, divided by the surface area of the Moon gives the value of 0.04/km$^2$, which will be the top right value on the thick black curve of Fig.~\ref{noCat}.

Similarly, the number of $D>1$~km craters produced by comets will be
$$
N_1=n(>1)/n(>75) N_{1/20}=4.2\times 10^6
$$
and the number of $D>12$~km asteroids hitting Mars is $8\times P^{ast}_{Mars/Moon} n(>12)/n(>10) =40$. The latter, divided by the surface area of Mars, gives $3\times 10^{-7}$/km$^2$,  which will be the top right value on the thick black curve of Fig.\ref{Mars}b.

\section{A reanalysis of the cataclysmic scenario} 
\label{Revisit}

\begin{figure}[t!]
\centerline{\includegraphics[width=9.cm]{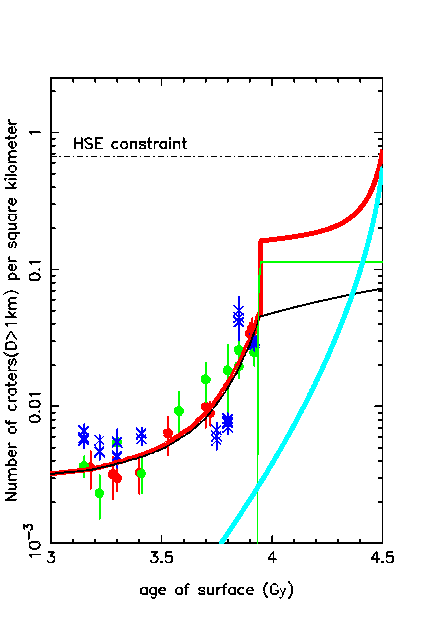}}
\caption{\footnotesize The number of craters produced by asteroids (thin black curve), comets (thin green curve) and leftover planetesimals (thick cyan curve ) as a function of age (i.e. the cumulative distribution $N_{crater}(<t)$). The red thick curve  is the sum of these three contributions, in a cumulative sense. {Because the time of the instability is unknown a priori, both the black and the green curves
can be shifted to the left or right arbitrarily.} The dots of various colors/symbols show the crater densities measured on units with well determined radiometric ages. See text for additional explanations. }
\label{miniCat}
\end{figure}

In this section we follow the analysis of M12, but using the results of the simulations presented in the previous section. The new result is presented in Fig.~\ref{miniCat}, which can be compared with Figs.~1 and 3b in the M12 paper.  

Fig.~\ref{miniCat} reports the density of craters (number per square kilometer) with diameter $D>1$~km as a function of lunar surface age, according to the crater counts in Neukum and Ivanov, (1984, green dots), Marchi et al. (2009, red dots) and  Robbins (2014, blue crosses) on terrains with well determined radiometric ages (none older than 3.92Gy). We report these three data sets, instead of just the Neukum and Ivanov data as in Fig. 1 of M12, in order to give an indication of the systematic uncertainties in crater counts, often exceeding the statistical errors, represented by the vertical bars. 

The black curve shows the crater densities produced by asteroids according to the simulations of Nesvorny et al. (2017), conducted and calibrated as explained in the previous section. However, a caveat needs to be stressed. Because 1 km craters are made by 50~m objects, we added the contribution of asteroids of this size continuously escaping from the main belt via the Yarkovsky effect (Morbidelli and Vokrouhlicky, 2003). {Following Neukum and Ivanov (1982), we assume that these objects produce a density of 1~km craters per km$^2$ growing as $8.38\times 10^{-4} t$, where $t$ is the age of the surface expressed in Gy}. In Fig.~\ref{miniCat} we added this function to that obtained from the numerical simulations of the asteroids destabilized by the giant planet instability.

The green curve in Fig.~\ref{miniCat} shows the crater densities produced by comets destabilized from the trans-Neptunian region, from the simulations of Nesvorny et al., 2013). 

Because the time of the instability is unknown a priori, both the black and the green curves can be shifted to the left or right arbitrarily. However, their vertical scale is fixed, from the considerations on the fluxes of  asteroids and comets, their size distribution and the scaling laws converting projectile size to crater size, all described in the previous section. 

The cyan curve in Fig.~\ref{miniCat} is the integral of the function $C_1(t)$ describing the decay of the number of craters as a function of age produced by planetesimals leftover from terrestrial planet formation (see previous section).  Because the number of these planetesimals still existing 4.5~Gy ago is unknown, what is relevant in the cyan curve is its shape, while the absolute (vertical) scaling is arbitrary.  

The red curve in Fig.~\ref{miniCat} is the sum of the asteroid, comet and planetesimal contributions in a cumulative sense. As discussed in M12 and in the Introduction section of this paper, the total number of impacts on the Moon (i.e. the value that the red curves reaches at $t=4.5$~Gy) has to correspond to the total mass accreted by the Moon since its formation, which is traditionally believed to be constrained by the HSE concentration in the lunar mantle. Assuming (i) that the accreted mass is $\sim 2.5\times 10^{-6} M_\oplus$) (Day et al., 2007; Day and Walker, 2015), (ii) that only $\sim 50$\% of the projectile mass is accreted by the Moon (Artemieva and Shuvalov, 2008) --the rest being lost by evaporation or debris-ejection and escape from the lunar gravitational well--, (iii) a projectile size-frequency distribution like that of main belt asteroids,(iv) an asteroid mean density of 2.6~g/cm$^3$ and (v) the projectile-crater scaling law of Holsapple and Housen (2007), M12 estimated that the total number of craters larger than 20~km in diameter per unit lunar surface ($N_{20}$) had to be $4.75\times 10^{-4}$. Thus, assuming the ratio $N_{1/20}=1400$ (Marchi et al., 2012), the density of craters larger than 1~km ($N_1$) has to be 0.65. {While each of the assumptions above could be debated (particularly (ii): the fraction of the colliding mass contaminating the mantle) the final result is unlikely to change by as much as an order of magnitude.} We report its value as an horizontal line in Fig.~\ref{miniCat}. In summary, the constraint on the HSEs lunar mantle concentration translates into the requirement that the red curve touches the horizontal dashed line at $t=4.5$~Gy. 

The fact that the red curve has a broken slope at $t\sim 3.95$~Gy reveals the existence of a cataclysm {(although one less pronounced than that originally envisioned by Tera et al., 1974, or Ryder, 2002)}. In fact, in a cumulative diagram like Fig.~\ref{miniCat} an impact surge translates into a broken curve, turning from steep to shallow as age progresses. This is because the impact rate is the derivative of the cumulative crater curve (red). It is evident from Fig.~\ref{miniCat} that the solution presented in the figure is the one consistent at the same time with the ``HSE constraint'' and the measured crater densities during the LHB. Shifting the black and green curves to the right or to the left (i.e. changing the timing of the cataclysm), or changing the vertical scaling of the cyan curve, would only make the fits worse. Thus, we confirm the result of Ryder (2002) and M12  that it is possible to reconcile the {bombardment $\sim 3.9$~Gy ago} with the small HSE-inferred accreted mass only if a cataclysm occurred. 

Comparing with the results of M12, however, we see that the cataclysm should have occurred later: at $\sim 3.95$~Gy instead of $\sim 4.1$~Gy. Moreover, the asteroidal contribution during the LHB, compared to the overall cratering of the Moon, would have been much smaller: less than 10\% of the total number of craters would have been produced by LHB asteroids, whereas this fraction was $\sim 1/4$ in M12. These differences would disappear if Nesvorny et al. (2017) had underestimated the asteroid bombardment by a factor $\sim 3$. However, as we will see in Sect.~\ref{Mars}, this is not possible because it would violate constraints on Mars.

Finally, in Fig.~\ref{miniCat} the comet contribution during the LHB (not considered in M12) dominates the asteroidal contribution by a factor of $\sim 2$. The predominance of cometary impacts during the LHB was already recognized in Gomes et al. (2005) and recently re-assessed in Rickman et al. (2017). Given the uncertainty related to the calibration of the comet flux on the population captured as Trojans of Jupiter, it is possible that the cometary contribution has been somewhat overestimated in Fig.~\ref{miniCat}. Nevertheless, even if possibly not dominating the LHB, the comets should have contributed to a substantial fraction of it. Note that, if comets smaller than $\sim 1$~km were less numerous than expected from an asteroidal SFD (Singer et al., 2016), they would have contributed much less to the formation of 1~km craters, but the black and green curves in Fig.~\ref{miniCat} would still give the correct relative contributions of asteroids and comets to basin-forming impacts.   

This conclusion on the importance of the cometary bombardment during the cataclysm may raise a problem for the cataclysmic scenario. In fact studies of platinum-group elements in lunar crustal samples from {roughly 4 Gy ago}, which presumably were delivered by the impactors, show the absence of primitive, carbonaceous chondritic material. This suggests that comets did not play a major role in the ancient bombardment (Kring and Cohen, 2002; Galenas et al., 2011). The same reasoning can be applied to the analysis of the projectile fragments in regolith breccias collected at the Apollo 16 site (Joy et al., 2012). We will see in section~\ref{New} that the problem of the cometary bombardment is nonexistent in the accretion tail scenario. 

If one postulates that comets disrupted into small pieces on their way into the inner Solar System, then the cataclysm in terms of crater-production, carried only by asteroids, would be very weak, with {only $4\times 10^{-2}$ km-size craters produced per square kilometer (equivalently, $\sim 4$ basins on the entire Moon, if an asteroid size frequency distribution is assumed)}. In this sense, we can speak of a ``mini-cataclysm''. Moreover, if comets are removed, the flux of leftover planetesimals should be increased relative to that shown in Fig.~\ref{miniCat}, in order to conserve the total mass delivered to the Moon. 

\section{The depletion of HSEs from the lunar mantle. What does the current HSE abundance really record?}
\label{Lunar}

The results of the previous section are based on the assumption that HSEs were removed from the lunar mantle only during metal-silicate fractionation when the lunar core formed, which was contemporary with the Moon's formation. Based on this assumption, the current HSE abundances in the lunar mantle would reflect the amount of chondritic material that the Moon accreted after its formation. However, HSEs can be removed from the mantle of a planet by the exsolution and segregation of iron sulfide (FeS) liquid well after core formation was complete (Rubie et al., 2016). The solubility of S in silicate liquid, termed the sulfur concentration at sulfide saturation (SCSS), decreases strongly with deceasing temperature (but also with increasing pressure) (Mavrogenes and O'Neill, 1999; Fortin et al., 2015; Wykes et al., 2015; Laurenz et al., 2016; Smythe et al., 2017). Thus, exsolution and segregation of FeS liquid occurs during the cooling and crystallization of planetary magma oceans as a consequence of the silicate liquid becoming supersaturated in S because of the temperature-dependence of SCSS. HSEs are removed together with the exsolved FeS because they are chalcophile and partition very strongly into FeS liquid compared to silicate liquid (Mungall and Brenan, 2014, Laurenz et al., 2016). As a result, the concentration of HSEs in Earth's mantle reflects the amount of chondritic material accreted subsequent to magma ocean crystallization, which may occur significantly later than the primary core-mantle differentiation that occurs during the formation of a planet (Rubie et al., 2016). 

Because of its relatively high density {(ca. 4.7 g/cm$^3$ at 4 GPa and 1800--2000 K compared with ca. 2.9 g/cm$^3$ for peridotite liquid at similar conditions)}, the segregation of exsolved FeS liquid can occur efficiently in a crystallizing magma ocean when the silicate melt fraction is high. However, as crystallization proceeds, segregation of FeS through the growing crystalline matrix actually becomes inhibited by the presence of silicate melt when the melt fraction is low (Holzheid et al., 2000; Rushmer and Petford, 2011; Holzheid, 2013; Cerantola et al., 2015). The critical value of the silicate melt fraction at which this happens is poorly known but is likely to be in the range 30-50\% (Stevenson, 1990; Minarik et al., 1996; Holzheid et al., 2000; Costa et al., 2009; Solomatov, 2015). Thus in the Earth, segregation of exsolved FeS liquid ended when the crystal/melt ratio of the magma ocean was 50-70\% and this point marked the start of late accretion and the addition of a late veneer. 

The results of Rubie et al. (2016) are developed in the framework of the early differentiation of the Earth, and their application to the Moon requires some caution. A major difference between the Moon and Earth is the very low average pressure in the lunar magma ocean, compared with that of the Earth's magma ocean, which has a large effect on SCSS. For the Moon we can consider a characteristic pressure of 1.5 GPa, which is the pressure at the depth of 375 km that separates the lunar mantle into two equal-mass layers. In contrast, for the Earth the equivalent pressure is of the order of 50 GPa. Because SCSS is strongly dependent on $P$, SCSS for the low pressure conditions of the lunar magma ocean is much higher (e.g. 2000-3000 ppm) than for Earth's mantle and is also significantly higher than the low sulfur concentration of $\sim 75$ ppm that has been estimated for the lunar mantle (Chen et al., 2015; {Hauri et al., 2015}; McCubbin et al., 2015). 

{The lunar magma ocean is considered to have crystallized from the bottom up (e.g. Elkins-Tanton et al., 2011). The silicate minerals (mainly olivine, pyroxene and plagioclase) that formed during magma ocean crystallization accommodate neither sulfur nor HSEs in their structures, at least not in significant concentrations, because these elements are highly incompatible. Thus in the Moon, sulfur and the HSEs became increasingly concentrated in the residual silicate magma that resided above the cumulative zone of crystallization and below the growing anorthositic crust. As magma ocean crystallization approached completion, the FeS concentration inevitably became high in the remaining low fraction of magma and eventually must have exceeded SCSS. When crystallization of the LMO reached completion, all sulfur would have exsolved as FeS liquid, irrespective of SCSS values. }

\begin{figure}[t!]
\centerline{\includegraphics[width=9.cm]{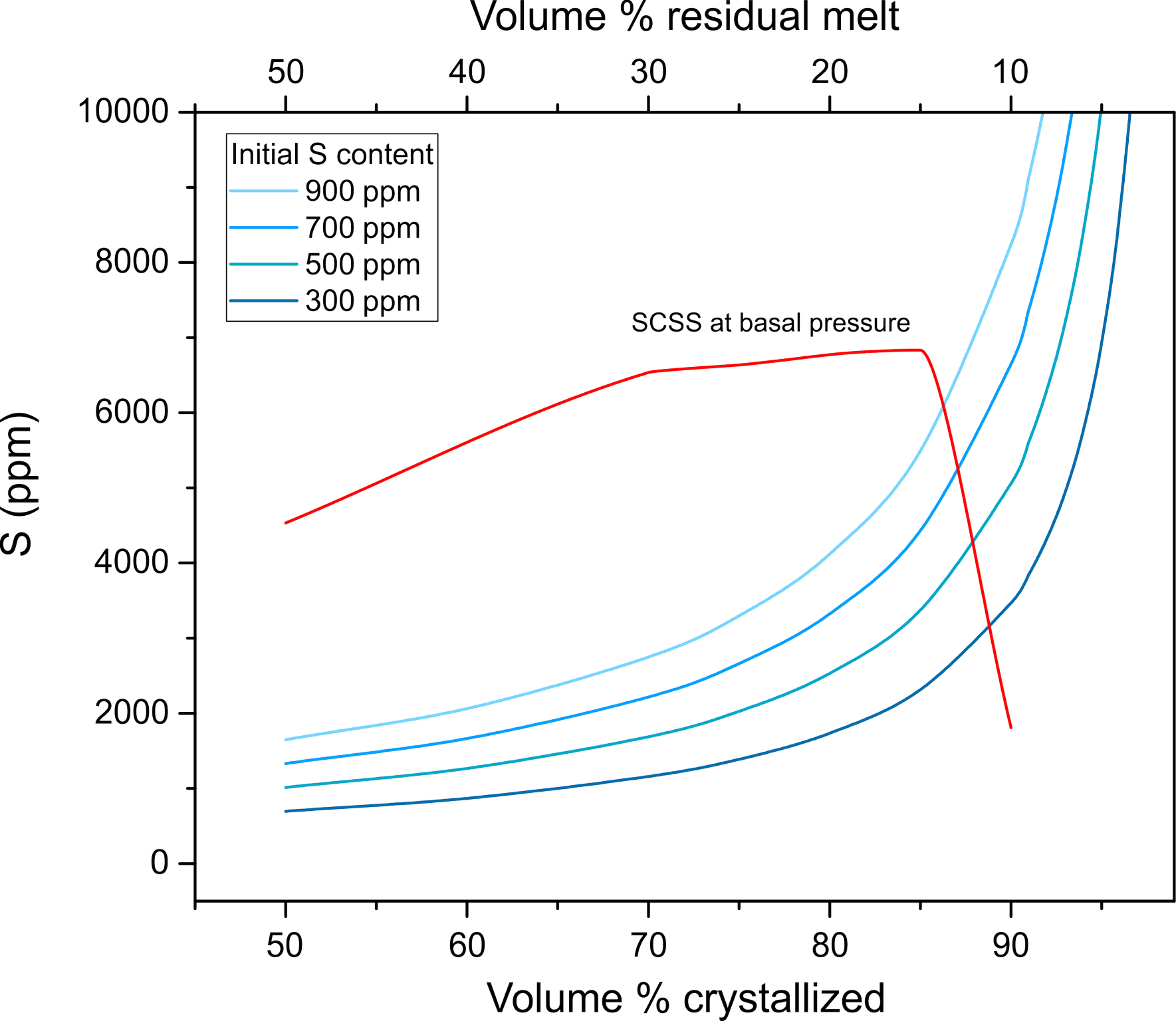}} 
\caption{\footnotesize {Blue curves show the sulfur concentration in the residual silicate melt layer of the LMO as a function of the volume percent of magma ocean crystallization and the equivalent volume percent of the residual melt. The four curves are for different initial bulk S contents in the range 300-900 ppm (at the time of the Moon's formation). The red curve shows SCSS calculated for the residual melt layer as crystallization proceeds, based on the evolving pressure at the base of the melt layer, according to the parameterization of Smythe et al. (2017). The strong decrease of SCSS after $\sim$85\% of the LMO has crystallized is caused by the evolving composition of the residual silicate liquid during the late stages of LMO crystallization (Elkins-Tanton et al., 2011). These results show that FeS exsolution starts after 85-90\% of the LMO has crystallized (where the red curve and the appropriate blue curve intersect). }}
\label{SCSS}
\end{figure}

{The evolution of SCSS during LMO crystallization determines the depth at which FeS exsolution commences. In order to calculate SCSS as a function of residual silicate melt fraction, we use the SCSS parameterization of Smythe et al. (2017) which is based on low-pressure experimental data and takes the effect of silicate liquid composition into account. The composition of the silicate melt evolves as magma ocean crystallization proceeds, and is calculated using the LMO crystallization model of Elkins-Tanton et al. (2011) over a range of residual melt fractions. For each residual silicate melt layer, we computed the basal pressure of the LMO. Temperature at a given pressure was fixed approximately mid-way between the liquidus and solidus (Rai and van Westrenen, 2014; Suckale et al. 2012). We calculate SCSS using the pressure at the base of each melt layer, where sulfides would start to exsolve because SCSS decreases with increasing pressure. Between residual melt fractions of 30 vol.\% and 10 vol.\%, SCSS decreases to around 1800 ppm (Fig.\ref{SCSS}), mainly due to the evolution of the residual silicate melt composition. }

{In order to calculate HSE evolution during lunar differentiation and magma ocean crystallization, we need to make an assumption about the Moon's bulk S content. One possibility is that the Moon inherited the S and HSE concentrations of Earth's mantle immediately prior to the Moon-forming giant impact. This assumption is not unreasonable given the isotopic and compositional similarities between the compositions of the terrestrial mantle and the Moon, which suggest their mutual equilibration (Pahlevan and Stevenson, 2007; Lock et al., 2016; Lock and Stewart, 2017). {Similarities in S concentration between the Earth and lunar mantle are also evident from the partitioning of trace elements (e.g. Steenstra et al. 2017).} The concentrations of S and HSEs in Earth's mantle prior to the Moon forming event are unknown but plausible values can be obtained from the model of Rubie et al. (2016), namely $\sim 300$ ppm S, 1.9 ppb Pt, 16 ppb Pd, 3.4 ppb Ru and 1.4 ppb Ir. However, as discussed below, based on estimates of the S content of the lunar core, the bulk S content of the Moon could be considerably higher (e.g. 900 ppm). We therefore consider a range of bulk S contents.}

{The evolution of the S concentration in residual melt during LMO crystallization is shown for a range of bulk lunar S concentrations in Fig.~\ref{SCSS}. Based on the initial S content, we first calculate the S contents of mantle and core, immediately following lunar core formation, by mass balance using metal-silicate partition coefficients for S as parameterized by Boujibar et al. (2014). Using the silicate melt composition of a fully molten LMO (Elkins-Tanton et al. 2011) and assuming that the conditions of core-mantle differentiation were 4.8 GPa and 2250$^\circ$C (Steenstra et al., 2016), the sulfur partition coefficient $D_S^{\rm metal-silicate} \sim 18$. Thus, with an initial S content for the Moon of 300 ppm, the sulfur concentration of the LMO is $\sim 240$ ppm immediately after core-formation, with $\sim 0.4$ wt\% S in the core, consistent with the results of Steenstra et al. (2016). Then, we assume that, after core formation, the Moon accreted 0.2\% of its mass (or $2.5\times 10^{-5} M_\oplus$) from chondritic (CI) material. This amount of mass, which is justified below in section 5, would have delivered $\sim 108$ ppm S, thus bringing the total S concentration in the LMO to $\sim 350$ ppm\footnote{We assume that no S loss occurred upon impact. The latter is supported by S isotopic compositions of mare basalts, indicating that less than 1-10\% S was lost after the Moon-forming impact (Wing and Farquhar 2015).}. In contrast, if we assume an initial S content of 900 ppm and perform the same set of calculations, the LMO would have contained 825 ppm S.}

{Fig.~\ref{SCSS} shows that the S concentration in the LMO exceeds SCSS when more than 85-90\% of the magma ocean had crystallized. At this point, FeS liquid starts to exsolve. For a higher initial S content of 900 ppm, FeS starts to exsolve when $\sim 80$\% of the magma ocean has crystallized.  In either case, exsolved FeS liquid would have been concentrated at the top of the lunar mantle, probably in association with the late crystallizing ilmenite-bearing cumulate layer that is also considered to have been enriched in KREEP. Because the HSEs are incompatible and strongly chalcophile, the entire mantle budget of these elements would have partitioned into the exsolved FeS liquid by the time LMO crystallization was complete. }

{Based on the same assumptions as for S (metal-silicate partitioning during core formation followed by subsequent addition of 0.2\% CI material), the HSE concentrations in the LMO are calculated to be 0.75 to 1.6 ppb, which is an order of magnitude greater than the current lunar mantle HSE concentrations of 0.1-0.2 ppb (Day et al., 2007). This estimate is independent of the initial composition of the Moon, because the HSEs are extracted entirely to the lunar core during core formation because their metal-silicate partition coefficients are extremely high, especially at the low pressures of the Moon's interior (Mann et al., 2012). Similarly to S, the HSEs would have been increasing enriched in the residual silicate melt layer during magma ocean crystallization, because they are highly incompatible. The HSEs partition strongly into the sulfide when FeS exsolves, thus resulting in HSE concentrations in the residual silicate liquid of $10^{-4}$ to $10^{-5}$ ppb. Essentially the entire budget of HSEs then resides in the exsolved FeS sulfides,  which subsequently governed the fate of the HSEs. }

{The density structure of the lunar mantle was unstable towards the end of magma ocean crystallization because the upper late-crystallizing mantle layer was enriched in Fe/Mg and contained late-crystallizing dense oxide phases such as FeTiO3 ilmenite (Hess and Parmentier, 1995; Solomatov, 2000; Elkins-Tanton et al., 2011). The unstable density structure caused mantle overturn to occur near the end of magma ocean crystallization that resulted in liquid FeS, crystalline ilmenite, and some quantity of KREEP to sink to the deep mantle because of their high density. Thus the exsolved FeS phase, containing virtually the entire mantle inventory of HSEs, was transported to deep levels in the mantle, leaving the upper mantle strongly depleted in HSEs ($10^{-4}$ to $10^{-5}$ ppb as shown above). Some authors consider that the high-density material sank all the way to the bottom of the mantle where it formed a global layer above the core (Zhong et al., 2000; Stegman et al., 2003;  Scheinberg et al., 2015). If this assemblage contained sufficient quantities of KREEP, the material would slowly heat up, and potentially rise buoyantly. At some stage, this material would melt, thus accounting for the high-titanium basalts. The latter provide evidence for an association between ilmenite and FeS because they contain higher concentrations of sulfur (1500-2700 ppm) than the low-titanium basalts (500-1500 ppm S) (Taylor, 1975; Danckwerth et al., 1979). {However, due to the incompatible behavior of S, fractional crystallization possibly contributed to the generation of high S contents in high-Ti mare basalts}.}

{A significant proportion of the exsolved FeS may have migrated to the lunar core following mantle overturn for at least two reasons. (1) There has to have been at least some decoupling of liquid FeS from solid ilmentite because Ti-rich basalts are generally considered to be undersaturated in sulfur. (2) Metal-silicate partitioning of sulfur during lunar core formation results in a low concentration of S in the core (e.g. 0.1-1.0 wt\%, depending on the Moon's bulk S content) as discussed above. On the other hand, the core contains likely 3-8 wt\% S, according to geophysical evidence based on seismic velocity and core density determinations (e.g., Weber et al., 2011; Garcia et al., 2011, Antonangeli et al., 2015). Such S contents can be achieved if some or all of the FeS liquid that was transported to the deep mantle by mantle overturn segregated from the ilmenite-bearing lithologies and migrated to the core. For example, if this separation process was 100\% efficient, $\sim 6$ wt\% S would result in the core if the bulk S content of the Moon (including S accreted after core formation but before mantle overturn) is 860 ppm. Such segregation is likely to have occurred during melting at the base of the mantle as a consequence of radiogenic heating caused by KREEP components. Without significant melting, it is unlikely that FeS liquid could segregate from the ilmenite/KREEP/FeS mixture because of its low volume fraction in the mantle rocks and wetting (dihedral) angles in olivine aggregates that significantly exceed 60$^\circ$ (Terasaki et al., 2008). }

{Subsequent to LMO crystallization and mantle overturn, a late veneer of CI chondrite composition (0.02\% of the lunar mass) was accreted to the Moon (Day et al., 2007; Day and Walker, 2015), in agreement with the accretion model presented in section~\ref{New}. This brings the final abundances in the lunar mantle to 0.17 ppb Pt, 0.13 ppb Ru, 0.08 ppb Ir and 0.12 ppb Pd. These calculated abundances of the HSEs are in good agreement with the estimated actual values of 0.1 ppb Pd, Ru and Ir, and 0.2 ppb Pt (Day et al., 2007). Most importantly, the final abundances of the various HSEs are in chondritic proportions relative to each other.} 
   
\subsection{When did the lunar magma ocean crystallize?}

{We have argued above that depletion of HSEs from the lunar mantle was caused by the exsolution of liquid FeS after 80-90\% of the mantle had crystallized. This was followed by mantle overturn, which transported the exsolved FeS together with the entire inventory of HSEs to deep regions of the mantle and possibly to the core. Therefore, for defining the time at which the retention of mantle HSEs started, the important question concerns the final crystallization age of the lunar magma ocean relative to the age of Moon formation.}

The crystallization of the magma ocean regulates the age  of  the  earliest lunar crust. The issue of lunar crust formation is a highly contentious one and contrasting ages can be found in the literature.  For instance Borg et al. (2011) argued for a young lunar crustal formation age (4.35 Gy, i.e. about 100--150~My after lunar formation) from the oldest ferroan anorthosite samples with concordant multiple radioactive chronometers. Yet, there are claims for much older ages in some samples ($\sim 4.44$~Gy; Norman et al., 2003).
 
From the modeling point of view it {appears} that, as long as the surface was molten, the Moon would have solidified 80\%  by volume in about 1~ky (Elkins-Tanton et al., 2011). However, the formation of floating plagioclase should  have  formed  a  thermally  conductive,  global  anorthosite lid  on  the  Moon, delaying to $\sim$10~My the final crystallization of the magma ocean in the last 100 km thick layer beneath nascent crust. And, if tidal heating is taken into account (Meyer et al., 2010; Chen and Nimmo, 2013), the ultimate crystallization could have been delayed by up to 100 or even 200~My (Elkins-Tanton et al., 2011). If this is the case, the HSE content of the upper lunar mantle would not record the (potentially large) quantity of material accreted during this long timespan. On the other hand, holes driven into the thermally conductive lid by early impacts could have favored a more rapid cooling of the underlying magma ocean (Perera et al., 2017). {One should also consider the possibility that the magma ocean could have completely crystallized first on the farside, and that the nearside crystallization was prolonged (because of the concentration of KREEP on the nearside).}

A new study on lunar zircon analyses (Barboni et al., 2017) claimed that the final crystallization of the lunar magma ocean was very early, i.e. about 4.51~Gy ago. Strictly speaking, however, this work shows that the lunar magma ocean fractionated to the point of zircon stability by 4.51 Gy ago. This may not be the final crystallization time. It is consistent with relatively rapid cooling to the point of zircon solidification, followed by gravitationally-driven overturn with attendant adiabatic melting. This new melt would rise to under the partially-formed anorthosite crust, with older zircons embedded. Tidal heating is expected to flex the anorthosite lid and retain a molten shell beneath it for as long as 200~My (Meyer et al., 2010). Thus, the data from Barboni et al. is unlikely to represent the final solidification age of the whole Moon. Extensive anorthosite (Carlson and Lugmair, 1988; Borg et al., 1999, 2011; Boyet et al., 2015), KREEP (Carlson and Lugmair, 1979; Nyquist and Shih, 1992; Taylor et al., 2009; Borg et al., 2015; Snape et al., 2016), and Mg suite ages (Boyet and Carlson, 2007; Gaffney and Borg, 2014; Barboni et al., 2017) indicate a long solidification history.

Conductive heat transfer constraints prevent the instantaneous formation of the anorthosite crust. It had to have formed over a period of time longer than the errors on radiogenically-dated samples; the question is, how much longer? There can be no single age for the lunar crust, and the chances that we sample either the first or the last crust is highly unlikely (and the most-recently formed crust is probably at depth).

\section{The LHB as a tail-end of planet accretion} 
\label{New}

On the basis of the results of the previous section, we now free ourselves from the constraint on the total mass accreted by the Moon since its formation, given that this quantity may not be tracked by the lunar HSE concentrations. 

\begin{figure}[t!]
\centerline{\includegraphics[width=9.cm]{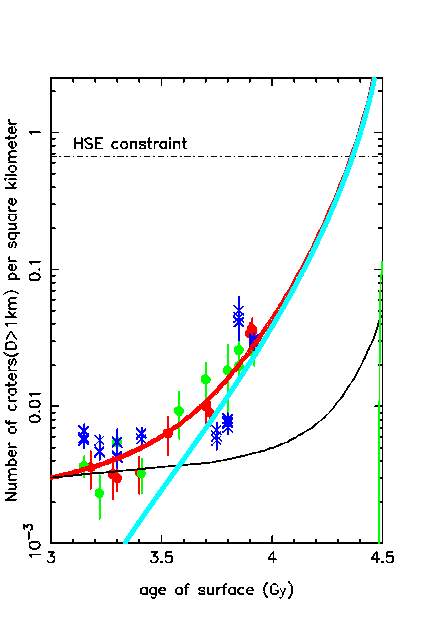}}
\caption{\footnotesize The same as Fig.~\ref{miniCat}, but for a giant planet instability 4.5~Gy ago, and a population of leftover planetesimals increased by a factor of 10. In this case the LHB crater data are well fit in an accretion tail scenario.}
\label{noCat}
\end{figure}

In this case, we can scale upwards the amount of material delivered to the Moon by leftover planetesimals and shift the time of the giant planet instability until we fit well the lunar crater data. The best result is shown in Fig.~\ref{noCat}. In this case there is no cataclysm: the red curve is smooth and the giant planet instability that triggers the asteroid and cometary bombardment has been set at 4.5~Gy. Nevertheless, the crater densities in the LHB period are well reproduced by the model. The lunar bombardment declined monotonically and was dominated by leftover planetesimals until $\sim 3.7$~Gy ago, unlike in the cataclysm model where the LHB was caused by an impact spike and dominated by comets and asteroids (section~\ref{Revisit} and Fig.~\ref{miniCat}). Thus, in this case, the LHB is explained in the accretion tail scenario. 

{The expected cometary spike at the time of the giant planet instability} is no longer a problem because it would have occurred very early in the lunar history, without leaving morphological or chemical traces. This would explain the lack of cometary signatures at the LHB time (Kring and Cohen, 2002; Galenas et al., 2011). 

In Fig.~\ref{noCat} we postulated that the total impacting mass was 10 times larger than in Fig.~\ref{miniCat}, i.e. it was $5\times 10^{-5} M_\oplus$. Like before, we assume that only 1/2 of this material is incorporated in the Moon, the rest being lost by escape of vaporized or solid projectile material from the lunar potential well  (Artemieva and Shuvalov, 2008). Thus, the total amount of HSEs accreted by the Moon corresponds to that contained in $2.5\times 10^{-5} M_\oplus$ of chondritic material. Consequently, to be consistent with the current HSE concentrations in the lunar mantle (the horizontal line in Fig.~\ref{noCat}), the HSEs should have been recorded only since $\sim 4.35$~Gy ago, about 100-150 My after lunar formation. We cannot fail noticing that this age corresponds to the preferred lunar crust age of Borg et al. (2011), although this statement has to come with all the caveats reported in the previous section. Given the uncertainties on the decay rate of the bombardment carried by leftover planetesimals (Fig.~\ref{newDecay}) an HSE retention age of 4.40~Gy is also acceptable. These ages argue for a relative late crystallization of the lunar magma ocean, and are well within the range of possibilities estimated in Elkins-Tanton et al. (2011). 

The comparison between Figs.~\ref{miniCat} and~\ref{noCat} shows that the lunar crater record alone is not sufficient to discriminate between the cataclysm and accretion tail scenarios. The crater record would need to be extended to surface units older than 4~Gy, but in the absence of new sample-return missions this is not possible. For this reason, in the next sections we turn our attention to other constraints, in order to achieve a better view of the comparative advantages and disadvantages of the two scenarios.

\section{Implications}
{We discuss here some important implications of the new scenario presented in sect.~\ref{New} on the interpretation of the differences in HSE concentrations and W-isotope composition between the Earth and the Moon.}

\subsection{Difference in HSE concentrations}
\label{Earth-Moon}

As described above, the concentration of HSEs in the upper lunar mantle, if uniform throughout the mantle, implies that the Moon accreted  $2.5\times 10^{-6} M_\oplus$ of chondritic material after the formation of its core (Day et al., 2007; Day and Walker, 2015). Instead, the concentration of HSEs in the terrestrial mantle, which is known to be rather uniform, implies that the Earth accreted about $5\times 10^{-3} M_\oplus$ (Walker, 2009). Thus, the ratio of accreted materials (a.k.a. late veneers) is about 2,000 in favor of the Earth. 

This large ratio is surprising, because the ratio of accretion cross-sections, once gravitational focusing is taken into account, is about 20. So, the question why the Moon accreted so little material compared to the Earth is a prominent one. 

Bottke et al. (2010) proposed a first solution. If the planetesimal mass distribution was dominated by the largest bodies, it is in principle possible that 99\% of the late veneer mass accreted by the Earth was carried by less than 20 bodies. In this case, the ratio of accretion rates between the Earth and the Moon (20) implies that the Moon likely accreted none of these bodies. If numerous small planetesimals delivered only 1\% of the terrestrial late veneer mass and the Moon accreted just 1/20 of those, the ratio in final accreted masses would be 2,000:1, as observed. This scenario has been supported from the dynamical point of view by Raymond et al. (2013), while Marchi et al. (2014) supported its statistical aspect with Monte Carlo simulations accounting for different realistic planetesimals size distributions. The problem is that delivering 99\% of the terrestrial late veneer to the Earth in less than 20 impacts requires that the impactors were significantly bigger than Ceres.  While there is no reason to limit the planetesimal size distribution to Ceres-size, it is likely that these large bodies were fully differentiated, thus with a substantial fraction of their HSEs sequestered in their cores. Therefore, the scenario of Bottke et al. can work only if the cores of these large projectiles completely dissolved, oxidized and remained in the terrestrial mantle.  {Recent SPH simulations suggest that for projectiles larger than 1,500 km in diameter only 20 to 50\% of the impactor's core material remains in the target's mantle (Marchi et al., 2017). This reduces the delivery of highly siderophile elements to the Earth’s mantle and imply a terrestrial late accretion mass two to five times greater than previously thought, making the unbalance with the lunar late accretion mass even more extreme}.  Finally, the requirement in Bottke et al. (2010) that only 1\% of the terrestrial late veneer has been delivered by sub-Ceres planetesimals translates into the requirement that the total mass of these planetesimals in the inner Solar System at the end of terrestrial planet formation was less than $10^{-3} M_\oplus$ (Brasser et al., 2016). This may be problematic in the context of planet formation, even including planetesimal collisional grinding (Walsh and Levison, 2016). Moreover, if the planetesimal population had really been so small, it is likely that the terrestrial planets would have remained on orbits with too large eccentricities and inclinations because of insufficient dynamical friction (Jacobson and Morbidelli, 2014).

Schlichting et al. (2012) proposed an alternative scenario in which the late veneer was carried by small planetesimals (less than 10~m in size) which formed a collisionally damped disk, like Saturn's rings. Due to the small velocity dispersion of this disk, the gravitational focusing ratio between the Earth and the Moon would have been highly enhanced, potentially explaining a large imbalance in late veneer masses. However, this scenario can explain late veneer ratios up to only $\sim 200$, i.e. ten times smaller than the measured ratio. More importantly, there is no evidence that such a collisionally damped disk of small planetesimals ever existed in the inner Solar System. 

The idea (Rubie et al., 2016) that HSEs are retained only after the crystallization of the magma ocean can easily explain the imbalance in HSEs concentrations between the Earth and the Moon. In fact, the magma ocean of the Earth would have crystallized in only a few My after the Moon-forming event (Elkins-Tanton, 2008), whereas that of the Moon could have done so $\sim 100$~My later (Elkins-Tanton et al., 2011), as advocated in the previous section. So, terrestrial HSEs would track the full amount of material accreted since the Moon forming event, whereas lunar HSEs would track only the material accreted since a later time. For instance, in the scenario illustrated in Fig.~\ref{noCat}, the mass of planetesimals hitting the Moon since 4.5~Gy is 1/100 of that hitting the Earth ($5\times10^{-5} M_\oplus$ vs. $5\times 10^{-3} M_\oplus$), but the final HSE concentration in the upper lunar mantle is the observed one if HSEs were retained only since $4.35$~Gy (possibly 4.40~Gy). 

A ratio of 1/100 is smaller than the expected ratio of 1/20 (the ratio of the accretional cross-sections of the Moon and the Earth), but one should take into account that, because of small number statistics combined with a larger collision probability, the Earth samples the large-size end of the projectile distribution better than the Moon (Bottke et al., 2010; Marchi et al., 2014), leading to some deviation in the accreted masses relative to the expected 1/20 ratio. A ratio of 1/100 can be achieved if the projectiles had an asteroid-like size distribution that extended barely beyond Ceres-size ($D=973$km).

\subsection{Difference in Tungsten isotope ratios}

{A potential} constraint on the ratio of late veneer masses between the Moon and the Earth is provided by their respective Tungsten isotope ratios. In fact, a difference in the Tungsten isotope composition between the Moon and the Earth has been measured by Touboul et al. (2015) ($0.20 \pm 0.05$ in $\epsilon ^{182}$W units) and Kruijer et al. (2015) ($0.27 \pm 0.04$ in $\epsilon ^{182}$W units). {There are two possible interpretations of this difference. The first interpretation is that, at the time of formation, the Earth and the Moon had identical isotopic compositions for all elements (including W), due to a rapid equilibration between the Earth and the proto-Lunar disk (Pahlevan and Stevenson, 2007; Lock et al., 2016; Lock and Stewart, 2017). In this case, the current difference in $\epsilon ^{182}$W would be due entirely to different amounts of late veneer masses accreted by the Earth and the Moon. This is the interpretation given by both Touboul et al. (2015) and Kruijer et al. (2015). The second interpretation is that the Earth and the projectile that gave origin to the Moon had identical isotopic compositions for all cosmogenic isotopes, because they formed in the same region of the Solar System (Dauphas, 2017). But the $^{182}$W$/^{184}$W was different because $^{182}$W is the daughter product of $^{182}$Hf and its final concentration in the mantle of a body depends on the timescale of differentiation, which is different for bodies of vastly different masses. If most of the mass of the Moon was inherited from the projectile, the difference in $\epsilon ^{182}$W between the Moon and the Earth is mostly primordial, only partially altered by the chondritic material accreted subsequently by the two bodies.}

It should be noted that W is lithophile in presence of {FeS (Li and Audetat, 2012, 2015) so, unlike the HSEs, it would not have been sequestered into the FeS droplets when the latter exolved during magma ocean crystallization. Thus, the value of $\epsilon ^{182}$W for the Moon would have been affected by the entire mass accreted by the Moon after its formation, and not just by the mass accreted since LMO crystallization, recorded by the HSE budget.}

\begin{figure}[t!]
\centerline{\includegraphics[width=8.cm]{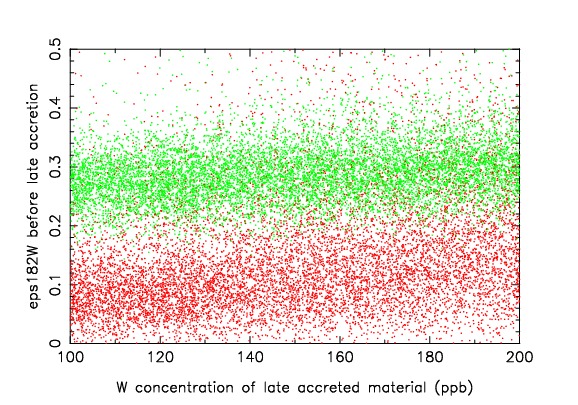}}
\centerline{\includegraphics[width=8.cm]{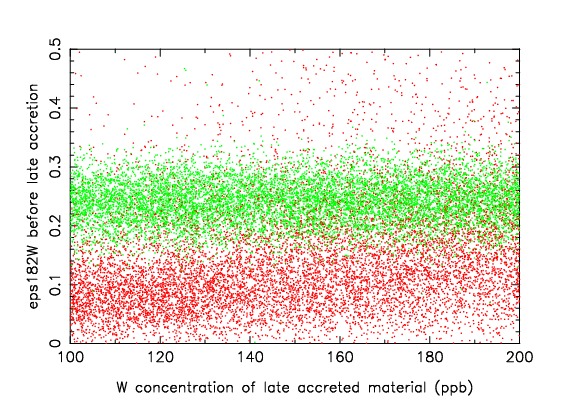}}
\caption{\footnotesize The value of $\epsilon ^{182}$W for the Earth (red) and the Moon (green) just after the Moon forming event, as a function of the W concentration in the late veneer material. These values are obtained by taking the current values of $\epsilon ^{182}$W of the Earth and the Moon and subtracting the contribution of the late veneers on these two bodies. The top panel refers to the scenario in which the Moon accreted a late veneer of $2.5\times 10^{-5} M_\oplus$, while the bottom panel refers to the classical scenario in which the Moon accreted only $2.5\times 10^{-6} M_\oplus$, as inferred from its HSE budget. Each pair of red-green dots refer to one trial in the Monte Carlo simulation, where the W concentration in the mantles of the Earth, of the Moon and in the late veneer material, as well as the late veneer mass on the Earth and the current $\epsilon ^{182}$W difference between the Moon and the Earth are drafted from the probability distributions described in the text. See online publication for a color version of this figure.}
\label{epsW}
\end{figure}

{We have performed a Monte Carlo simulation, in which we start from the current difference in  $\epsilon ^{182}$W between the Earth and the Moon and we substract the contribution of the late veneer masses acquired by the two bodies. The goal is to compute the original, post-formation values of $\epsilon ^{182}$W of the Earth and the Moon and verify which of the two interpretations discussed above is the most appropriate for our scenario, in which the Moon accreted a mass ten times larger than usually assumed. The Monte Carlo technique is used because of the uncertainties on several parameters.  We assume that the late veneer mass of the Earth was $(5.5\pm 2.5)\times 10^{-3} M_\oplus$ (1$\sigma$ uncertainty), whereas we fix the late veneer mass of the Moon to $2.5\times 10^{-5} M_\oplus$, as required in our scenario for the lunar bombardment. Moreover we assume that the W concentration in the mantles of the Earth and the Moon is $(13\pm 5)\times 10^{-3}$~ppb (1$\sigma$ uncertainty; the two concentrations are drafted independently) and that the average W concentration in the late veneer material was between 100 and 200~ppb (with a flat probability distribution). The $\epsilon ^{182}$W value of the late veneer material is assumed to be chondritic, i.e. $-0.19$. Finally, we consider that the current difference in $\epsilon ^{182}$W between the Moon and the Earth is $(0.235\pm 0.075)$ (2$\sigma$ uncertainty) that we obtain by combining the results of Touboul et al. (2015) and Kruijer et al. (2015). The result is shown in the top panel of Fig.~\ref{epsW}. The red dots show the original $\epsilon ^{182}$W of the Earth; the green dots that of the Moon. Each pair of red-green dots represent one Monte Carlo trial with parameters chosen according to the distributions just described. The result is that, on average, the Moon and the Earth had a difference of 0.15-0.20 in $\epsilon ^{182}$W prior to the late veneer, which supports the second interpretation described above. However, there is a significant overlapping between the red and green dots, so also the first interpretation is compatible with our scenario. 

For reference, we have repeated the Monte Carlo simulation using the same assumptions, except for the late veneer mass of the Moon, now set to $2.5\times 10^{-6} M_\oplus$, as inferred from the HSE lunar budget. The result is shown in the bottom panel of Fig.~\ref{epsW}. The conclusions are basically the same. The Moon and the Earth are on average closer to each other, but they still have a positive difference on average, of about 0.10-0.15 in $\epsilon ^{182}$W units.}

\section{Mars}
\label{Mars}

We have computed the number of craters produced on Mars as a function of age in the cataclysm and accretion tail scenarios. The ratios of impact rates on Mars vs. the Moon for asteroids, comets and leftover planetesimals are those reported in Sect.~\ref{NewSims}. As a reference crater size we chose $D>150$~km, so that there is a clear constraint on the density of these craters on the southern hemisphere of the planet, without having to correct for crater saturation, erosion etc. As said in Sect.~\ref{NewSims} craters of this size on Mars are produced by $D>12$~km asteroids and $D>18$~km comets. The number of projectiles are scaled as a function of size according to the numbers reported in Table~I. 

\begin{figure}[t!]
\centerline{\includegraphics[width=9.cm]{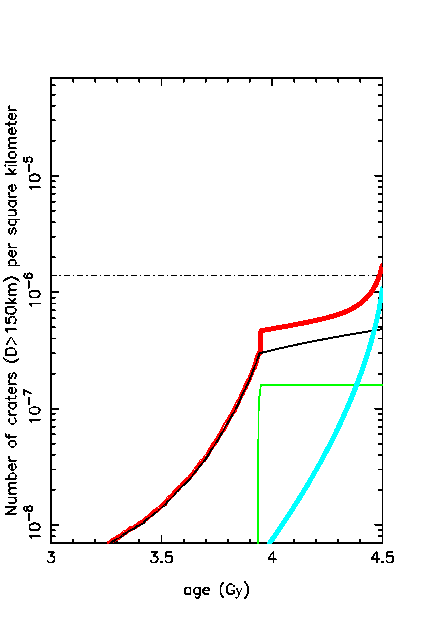}\,\includegraphics[width=9.cm]{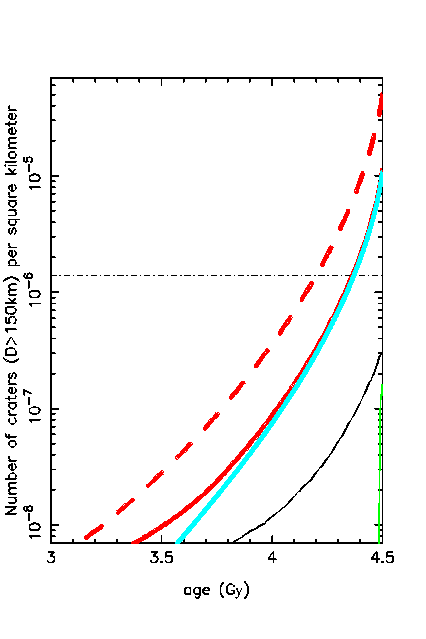} }
\caption{\footnotesize Density of craters (number per square kilometer) with diameter $D>150$~km as a function of Martian surface age in the cataclysm scenario (left panel) and in the accretion tail scenario (right panel). The meaning of the curves is the same as in Figs.~\ref{miniCat} and~\ref{noCat}. The horizontal thin line shows the surface density of $D>150$~km craters in the southern hemisphere of the planet. For reference, the dashed red curve in the right panel shows the density of $D>150$~km craters expected on the Moon in the accretion tail scenario. }
\label{FigMars}
\end{figure}

The results are illustrated in Fig.~\ref{FigMars}, in the left panel for the cataclysm scenario and in the right panel for the accretion tail scenario. The color codes of the different curves are the same as in Figs.~\ref{miniCat} and~\ref{noCat}.  

In the cataclysm scenario, the contribution of asteroids in the production of craters is about 1/3 of the total. Comets produce about half the number of craters of asteroids during the cataclysm, in contrast with the Moon, where they possibly dominate the crater production during the impact spike. The reason is that the impact rate ratio between Mars and the Moon is smaller for comets than for asteroids (Sect.~\ref{NewSims}). The total surface density of $D>150$~km craters produced in the last 4.5~Gy is consistent with the observed one. However, if Mars formed and solidified in just a few My (Dauphas and Pourmand, 2011; Elkins-Tanton, 2008) the cyan and red curves should be extrapolated backwards in time for about $\sim 50$~My, exceeding the observed limit by a factor of 2--3. 

Fig.~\ref{FigMars}a excludes the hypothesis of a cataclysm with an asteroidal flux 3 times larger than estimated in Nesvorny et al. (2017). {This enhanced flux would be needed if one wants to shift the beginning of the cataclysm to 4.1~Gy, while remaining in agreement with the lunar crater record (see Sect.~\ref{Revisit}). In this case, however, Fig.~\ref{FigMars}a shows} that the observed density of 150~km craters on Mars would be exceeded around 4.1--4.2~Gy.  Similarly, the figure can also exclude the scaling law for Mars impacts advocated by Bottke et al. (2016), because in this case the projectiles excavating a 150~km crater on Mars would be only 8~km in diameter and would be more than twice the number of 12~km asteroids.  So, the observed density of craters would be exceeded at 4.35~Gy. The problem of an excessive production of craters could be solved if Mars underwent a global resurfacing event at the corresponding time. The formation of Borealis could be such an event, if the latter is a giant basin due to the impact of a projectile larger than Ceres (Marinova et al., 2008; Nimmo et al., 2008; Citron et al., 2015). However, in the cataclysm scenario, the bombardment of Mars by asteroids and leftover planetesimals is too low to make such a collision plausible. Assuming a size distribution of projectiles like that of main belt asteroids, the curves shown in Fig.~\ref{FigMars}a imply that the probability of collision of a Ceres-size projectile with Mars would have been less than $\sim 3$\% in the last 4.5~Gy.    

In the accretion tail scenario (Fig.~\ref{FigMars}b), the production of craters on Mars is dominated by leftover planetesimals. The observed density of craters is largely exceeded which, at first sight, seems to provide a strong argument against this scenario. In this case, however, the hypothesis of a late formation of Borealis is more likely. Given the total cumulative bombardment and assuming an asteroid-like size distribution for the leftover planetesimals, the probability that a Ceres-size projectile hit Mars in the last 4.5~Gy is 25\%. However, slight differences in the size distribution of leftover planetesimals relative to asteroids can easily bring this probability close to unity. According to the red solid curve in Fig.~\ref{FigMars}b, Borealis formation (i.e. the global resurfacing event) should have occurred at 4.37~Gy (although different decay curves, such as that depicted by the red line in Fig.~\ref{newDecay}, would push this event back to $\sim 4.42$~Gy). Once we accept that a Borealis-forming impact happened on Mars, in the last 4.5~Gy the bombardment timeline depicted by the red solid curve in Fig.~\ref{FigMars}b implies that the probability that this event occurred less than 4.37Gy ago is 15\%.

An advantage of the accretion tail scenario over the cataclysm scenario is that in the former the total accreted mass $\sim 10^{-4} M_\oplus$ is roughly consistent with that deduced from the HSEs abundances in the Martian mantle (Walker, 2009), the exact value depending on when exactly Mars started to record HSEs. In the cataclysm scenario, instead, the total mass delivered during the integrated bombardment is insufficient by about an order of magnitude, and therefore the delivery of martian HSEs would require a singular event, such as the collision with a planetary embryo. 

The estimates of Martian surface ages based on crater counts typically assume that the production of craters of a given size per unit surface is the same as on the Moon (Neukum and Ivanov, 1994; Ivanov, 2001). The comparison between the cratering curves of Mars (solid red) and the Moon (dashed red) shown in Fig.~\ref{FigMars}b for the accretion tail scenario shows that this assumption leads to a severe underestimate of the age of the old Martian surfaces. This is because leftover planetesimals produce 4 times fewer craters of a given size per unit surface area on the red planet than on the Moon (Sect.~\ref{NewSims}), unlike asteroids which indeed produce roughly the same number of craters. In particular, in our model the boundary between the pre-Noachian and the Noachian eras would shift about 200~My back in time. Thus, the intense water activity on Mars in the Noachian could become easier to explain because a thick atmosphere is more likely to have existed at earlier times.

\section{Discussion}
\label{conclusions}

In this paper we have revisited the timeline of the lunar bombardment in the first $\sim$Gy of Solar System history. There are two contrasting views on the evolution of the bombardment in this period. One is the cataclysm scenario, in which the heavy bombardment that affected the Moon about 3.9~Gy ago (LHB) was due to a sudden surge in the impact rate (Tera et al., 1974; Ryder 2002). The other is the accretion tail scenario, in which the LHB was simply the tail-end of a more intense bombardment that declined over time since the phase of formation of the terrestrial planets (Hartmann, 1975; Neukum and Ivanov, 1994). Currently, there is no general consensus on what the actual evolution of the bombardment was. 

Like Ryder (2002) and Morbidelli et al. (2012) we consider that an important constraint is provided by the abundance of HSEs in the lunar (and martian) mantle(s). Under the traditional assumption that HSEs track the total amount of mass accreted by a planet after its core-mantle differentiation (presumably coeval with the main accretion phase of the body), we confirm the earlier results that the LHB would have to have been the consequence of a cataclysm. In fact, an accretion tail providing a sufficiently intense bombardment at $\sim 3.9$~Gy would have delivered about 10 times the mass constrained by the HSEs abundances in the lunar mantle.  Using new simulations from, mainly, Nesvorny et al. (2013, 2017), we find that the impact surge should have started about 3.95~Gy ago. During the cataclysm, comets would have had a large (possibly predominant) share of the impact rate on the Moon, while the bombardment of Mars would have been mainly asteroidal. 

However, a new result (Rubie et al., 2016) argues that HSEs are sequestered into a planet's core by FeS exsolution during the crystallization of the magma ocean, which can postdate the phase of metal-silicate segregation. We have shown in this paper that the same argument applies to the Moon, although with important differences relative to the Earth due to the low pressures of the lunar upper mantle. Moreover, unlike the Earth and Mars (Elkins-Tanton, 2008), the crystallization of the magma ocean on the Moon may have been a very slow process, taking up to 100-200~My to complete, due to the formation of an insulating anorthositic lid and internal heating from tides (Elikins-Tanton et al., 2011). In this case, the lunar HSEs would track only the material accreted from a time significantly delayed relative to the Moon formation event. 

We have shown (Fig.~\ref{noCat}) that in this case the LHB can be explained by an accretion tail, with no need for a late surge in the bombardment. If the HSEs have been recorded by the Moon only since 4.35--4.40~Gy, their final concentration in the upper lunar mantle would be consistent with the measurements (Day et al., 2007; Day and Walker, 2015).

Thus, we are back to the dilemma between the cataclysm and accretion tail scenarios. 

The accretion tail scenario has a number of advantages, already discussed in the body of this paper. There would be no prominent cometary bombardment during the LHB, consistent with the lack of corresponding chemical signatures in the lunar samples of the time (Kring and Cohen, 2002; Galenas et al., 2011). The delayed start of retention of HSEs in the lunar mantle, implied by the accretion tail scenario, would also explain in a simple way why the Moon is much more deficient in HSEs than the Earth, compared to their accretion cross-sections (a factor of 2,000 compared to a factor of 20; Walker, 2009). The total bombardment suffered by Mars in the accretion tail scenario would also be consistent with its HSE content.

There are additional advantages in the accretion tail scenario. It does not require that the instability of the giant planets occurred a long time after the removal of the protoplanetary disk (Gomes et al., 2005). An early instability is indeed the most generic outcome of dynamical models (Nesvorny and Morbidelli, 2012), whereas a delayed instability requires fine-tuning in the parameters of the trans-Neptunian disk (mass, distance from Neptune, dust production etc.; Deienno et al., 2017). An early instability and dispersion of the trans-Neptunian disk also explains why the most cratered surfaces on Pluto and Charon are consistent with 4 Gy of impacts in the current environment of the outer solar system (Greenstreet et al., 2015; Singer et al., 2016). If Pluto and Charon had remained embedded for a long time in a massive trans-Neptunian disk, as in the late instability hypothesis, it is likely that some portions of their surfaces would be more heavily cratered.  

On the flip side, the accretion tail scenario has some difficulties that need to be debated. First, while the cataclysm model gives a straightforward interpretation of the spike in the impact age distributions (Tera et al., 1974; Marchi et al., 2013), the accretion tail scenario has to concede that the existing impact records are biased towards younger impact events (Hartmann et al., 1975; {Haskin et al., 1998}; Norman, 2009; Boehnke and Harrison, 2016). We note however that the putative impact spike recorded in HED meteorites at $\sim 4.1$~Gy (Marchi et al., 2013) would not chronologically correspond with the beginning of the cataclysm, that our revised analysis in Sect.~\ref{Revisit} places at $\sim3.95$~Gy. 

Second, and perhaps most importantly, in the accretion tail scenario both the Moon and Mars would have been bombarded much more than what is revealed by their respective crater records.  Neumann et al. (2017) found a total of 40--50 lunar basins ($D>300$ km) in the analysis of GRAIL and LOLA data, whereas the accretion tail scenario would predict a number of basin-forming events in the last 4.5~Gy about 10 times larger. Similarly, on Mars, Fig.~\ref{FigMars} predicts a number of $D>150$~km almost 10 times larger than that recorded per unit surface in the southern hemisphere of the planet. Thus, reconciling the accretion tail scenario with the crater records requires that the topographic and gravitational signatures of big craters and basins could be retained only from $\sim 4.4$~Gy ago. For Mars, we have argued in Sect.~\ref{Mars} that this could be due to the formation of the Borealis basin, as a global resurfacing event. For the Moon, the explanation is more subtle. The oldest recorded basins on the Moon are very degraded, revealing that substantial viscous lateral flow in the crust occurred after their formation (Baldwin, 2006). Kamata et al. (2015) concluded that these basins should have formed in the first $\sim 50$~My after the crystallization of the lunar magma ocean. Thus, it is likely that basins formed on the anorthosite crust when there was still a magma ocean underneath could not be preserved due to the inability of the crust to support large topographic signatures at that time. 

In summary, given the currently available data, models and knowledge, our preference goes to the accretion tail scenario. Fortunately, this scenario implies two strong predictions on the age of Borealis formation and the crystallization age of the lunar magma ocean, which will be possible to test precisely in the future.

\acknowledgments 

We acknowledge support by the French ANR, project number ANR-13--13-BS05-0003-01  project MOJO (Modeling the Origin of JOvian planets), and the European Research Council (ERC) Advanced Grant ACCRETE (contract number 290568).

\section{References}

\begin{itemize}

\item[--] Antonangeli, D., Morard, G., Schmerr, N. C., Komabayashi, T., Krisch, M., Fiquet, G. and Fei, Y., 2015. Toward a mineral physics reference model for the Moon's core. PNAS, 112, 3916–3919. 
\item[--] Artemieva, N.A., Shuvalov, V.V. 2008. Numerical simulation of high-velocity impact ejecta following falls of comets and asteroids onto the Moon. Solar System Research 42, 329-334
\item[--] Baldwin, R.~B.\ 2006.\ Was there ever a Terminal Lunar Cataclysm?. With lunar viscosity arguments.\ Icarus 184, 308-318. 
\item[--] Barboni, M., Boehnke, P., Keller, B., Kohl, I.E, Schoene, B., Young, E.D, McKeegan, K.D., 2016.  Early formation of the Moon 4.51 billion years ago. Science Adv., Vol. 3, no. 1, e1602365
\item[--] Boehnke, P., Harrison, T.M. 2016. Illusory Late Heavy Bombardment. Proceedings of the National Academy of Science 113, 10802-10806. 
\item[--] Borg, L., Norman, M., Nyquist, L., Bogard, D., Snyder, G., Taylor, L., Lindstrom, M.\ 1999.\ Isotopic studies of ferroan anorthosite 62236: a young lunar crustal rock from a light rare-earth-element-depleted source.\ Geochimica et Cosmochimica Acta 63, 2679-2691. 
\item[--] Borg, L.E., Connelly, J.N., Boyet, M., Carlson, R.W. 2011. Chronological evidence that the Moon is either young or did not have a global magma ocean. Nature 477, 70-72. 
\item[--] Borg, L.~E., Gaffney, A.~M., Shearer, C.~K.\ 2015.\ A review of lunar chronology revealing a preponderance of 4.34-4.37 Ga ages.\ Meteoritics and Planetary Science 50, 715-732. 
\item[--] Bottke, W.F., Durda, D.D., Nesvorny, D., Jedicke, R., Morbidelli, A., Vokrouhlicky, D., Levison, H. 2005. The fossilized size distribution of the main asteroid belt. Icarus 175, 111-140. 
\item[--] Bottke, W.F., Walker, R.J., Day, J.M.D., Nesvorny, D., Elkins-Tanton, L. 2010. Stochastic Late Accretion to Earth, the Moon, and Mars. Science 330, 1527. 
\item[--] Bottke, W.F., Vokrouhlicky, D., Minton, D., Nesvorny, D., Morbidelli, A., Brasser, R., Simonson, B., Levison, H.F. 2012. An Archaean heavy bombardment from a destabilized extension of the asteroid belt. Nature 485, 78-81.
\item[--] Bottke, W.~F., Vokrouhlicky, D., Ghent, B., Mazrouei, S., Robbins, S., Marchi, S.\ 2016.\ On Asteroid Impacts, Crater Scaling Laws, and a Proposed Younger Surface Age for Venus.\ Lunar and Planetary Science Conference 47, 2036. 
\item[--] Boujibar, A., Andrault, D., Bouhifd, M.~A., Bolfan-Casanova, N., Devidal, J.-L., Trcera, N.\ 2014.\ Metal-silicate partitioning of sulphur, new experimental and thermodynamic constraints on planetary accretion.\ Earth and Planetary Science Letters 391, 42-54. 
\item[--] Boyet, M., Carlson, R.~W.\ 2007.\ A highly depleted moon or a non-magma ocean origin for the lunar crust?.\ Earth and Planetary Science Letters 262, 505-516. 
\item[--] Boyet, M., Carlson, R.~W., Borg, L.~E., Horan, M.\ 2015.\ Sm-Nd systematics of lunar ferroan anorthositic suite rocks: Constraints on lunar crust formation.\ Geochimica et Cosmochimica Acta 148, 203-218. 
\item[--] Brasser, R., Morbidelli, A. 2013. Oort cloud and Scattered Disc formation during a late dynamical instability in the Solar System. Icarus 225, 40-49.
\item[--]  Brasser, R., Mojzsis, S.J., Werner, S.C., Matsumura, S., Ida, S. 2016.  Late veneer and late accretion to the terrestrial planets. Earth and Planetary Science Letters. 455, 85-93. 
\item[--] Carlson, R.~W., Lugmair, G.~W.\ 1979.\ Sm-Nd constraints on early lunar differentiation and the evolution of KREEP.\ Earth and Planetary Science Letters 45, 123-132. 
\item[--] Carlson, R.~W., Lugmair, G.~W.\ 1988.\ The age of ferroan anorthosite 60025 - Oldest crust on a young moon?.\ Earth and Planetary Science Letters 90, 119-130. 
\item[--] Cerantola, F., Walte, N., Rubie, D.C. 2015. Deformation of a crystalline olivine aggregate containing two immiscible liquids: Implications for early core-mantle differentiation. Earth Planet. Sci. Lett. 417, 67-77.
\item[--] Chapman, C.~R., Cohen, B.~A., Grinspoon, D.~H.\ 2007.\ What are the real constraints on the existence and magnitude of the late heavy bombardment?.\ Icarus 189, 233-245. 
\item[--] Chen Y., Zhang, Y., Liu Y., Guan Y., Eiler J. and Stolper E. M. 2015. Water, fluorine, and 
sulfur concentrations in the lunar mantle. Earth Planet. Sci. Lett. 427, 37-46.
\item[--] Citron, R.~I., Genda, H., Ida, S.\ 2015.\ Formation of Phobos and Deimos via a giant impact.\ Icarus 252, 334-338.
\item[--] Cohen, B.~A., Swindle, T.~D., Kring, D.~A.\ 2000.\ Support for the Lunar Cataclysm Hypothesis from Lunar Meteorite Impact Melt Ages.\ Science 290, 1754-1756.  
\item[--] Costa, A., Caricchi, L., Bagdassarov, N. (2009) A model for the rheology of particle-bearing suspensions and partially molten rocks. Geochem. Geophys. Geosys 10, 3010.
\item[--] Danckwerth, P.~A., Hess, P.~C., Rutherford, M.~J.\ 1979.\ The solubility of sulfur in high-TiO2 mare basalts.\ Lunar and Planetary Science Conference Proceedings 10, 517-530. 
\item[--] Dauphas, N., Pourmand, A.\ 2011.\ Hf-W-Th evidence for rapid growth of Mars and its status as a planetary embryo.\ Nature 473, 489-492. 
\item[--] Dauphas, N.\ 2017.\ The isotopic nature of the Earth's accreting material through time.\ Nature 541, 521-524. 
\item[--] Day, J.M.D., Pearson, D.G., Taylor, L.A. 2007. Highly Siderophile Element Constraints on Accretion and Differentiation of the Earth-Moon System. Science 315, 217.
\item[--] Day, J.~M.~D., Walker, R.~J.\ 2015.\ Highly siderophile element depletion in the Moon.\ Earth and Planetary Science Letters 423, 114-124. 
\item[--] Deienno, R., Morbidelli, A., Gomes, R.~S., Nesvorny, D.\ 2017.\ Constraining the giant planets' initial configuration from their evolution: implications for the timing of the planetary instability. 2017. Astron. J. 153 153 
\item[--] Elkins-Tanton, L.~T.\ 2008.\ Linked magma ocean solidification and atmospheric growth for Earth and Mars.\ Earth and Planetary Science Letters 271, 181-191. 
\item[--] Elkins-Tanton, L.T., Burgess, S., Yin, Q.-Z. 2011. The lunar magma ocean: Reconciling the solidification process with lunar petrology and geochronology. Earth and Planetary Science Letters 304, 326-336. 
\item[--] Fernandez, Y.R., Jewitt, D., Ziffer, J.E. 2009. Albedos of Small Jovian Trojans. The Astronomical Journal 138, 240-250. 
\item[--] Fortin, M.-A., Riddle, J., Desjardins-Langlais, Y., Baker, D.~R.\ 2015.\ The effect of water on the sulfur concentration at sulfide saturation (SCSS) in natural melts.\ Geochimica et Cosmochimica Acta 160, 100-116. 
\item[--] Gaffney, A.~M., Borg, L.~E.\ 2014.\ A young solidification age for the lunar magma ocean.\ Geochimica et Cosmochimica Acta 140, 227-240. 
\item[--] Galenas, M.~G., Gerasimenko, I., James, O.~B., Puchtel, I.~S., Walker, R.~J.\ 2011.\ Continued Study of Highly Siderophile Element Characteristics of Apollo 17 Impact Melt Breccias.\ Lunar and Planetary Science Conference 42, 1413. 
\item[--] Garcia, R.F., Gagnepain-Beyneix, J., Chevrot, S., Lognonné, P. 2011.
Very preliminary reference Moon model. Phys. Earth Planet. Int. 188, 96-113
\item[--] Gomes, R., Levison, H.F., Tsiganis, K., Morbidelli, A. 2005. Origin of the cataclysmic Late Heavy Bombardment period of the terrestrial planets. Nature 435, 466-469.
\item[--] Greenstreet, S., Gladman, B., McKinnon, W.~B.\ 2015.\ Impact and cratering rates onto Pluto.\ Icarus 258, 267-288. 
\item[--] Hartmann WK (1975) Lunar “cataclysm”: A misconception? Icarus 24(2):181–187.
\item[--] Hartmann, W.~K., Ryder, G., Dones, L., Grinspoon, D.\ 2000.\ The Time-Dependent Intense Bombardment of the Primordial Earth/Moon System.\ Origin of the Earth and Moon 493-512. 
\item[--] Hartmann, W.~K.\ 2003.\ Megaregolith evolution and cratering cataclysm models--Lunar cataclysm as a misconception (28 years later).\ Meteoritics and Planetary Science 38, 579-593.
\item[--] Haskin, L.~A., Korotev, R.~L., Rockow, K.~M., Jolliff, B.~L.\ 1998.\ The case for an Imbrium origin of the Apollo Th-rich impact-melt breccias.\ Meteoritics and Planetary Science 33, 959-975. 
\item[--] Haskin, L.~A., Moss, W.~E., McKinnon, W.~B.\ 2003.\ On estimating contributions of basin ejecta to regolith deposits at lunar sites.\ Meteoritics and Planetary Science 38, 13-34. 
\item[--] Hauri, E.~H., Saal, A.~E., Rutherford, M.~J., Van Orman, J.~A.\ 2015.\ Water in the Moon's interior: Truth and consequences.\ Earth and Planetary Science Letters 409, 252-264.
\item[--] Hess, P.~C., Parmentier, E.~M.\ 1995.\ A model for the thermal and chemical evolution of the Moon's interior: implications for the onset of mare volcanism.\ Earth and Planetary Science Letters 134, 501-514.   
\item[--] Holsapple, K.~A., Housen, K.~R.\ 2007.\ A crater and its ejecta: An interpretation of Deep Impact.\ Icarus 187, 345-356.
\item[--] Holzheid, A. 2013. Sulphide melt distribution in partially molten silicate aggregates: implications to core formation scenarios in terrestrial planets. Eur. J. Mineral. 25, 267–277. 
\item[--] Holzheid A., Schmitz M. D., and Grove T. L. 2000. Textural equilibria of iron sulphide liquids in partly molten silicate aggregates and their relevance to core formation scenarios. J. Geophys. Res. 105, 13555-13567.
\item[--] Ivanov, B.~A.\ 2001.\ Mars/Moon Cratering Rate Ratio Estimates.\ Space Science Reviews 96, 87-104. 
\item[--] Jacobson, S.A., Morbidelli, A., Raymond, S.N., O'Brien, D.P., Walsh, K.J., Rubie, D.C. 2014. Highly siderophile elements in Earth's mantle as a clock for the Moon-forming impact. Nature 508, 84-87.  
\item[--] Jacobson, S.~A., Morbidelli, A.\ 2014.\ Lunar and terrestrial planet formation in the Grand Tack scenario.\ Philosophical Transactions of the Royal Society of London Series A 372, 0174. 
\item[--] Johnson, B.C., Collins, G.S., Minton, D.A., Bowling, T.J., Simonson, B.M., Zuber, M.T. 2016. Spherule layers, crater scaling laws, and the population of ancient terrestrial impactors. Icarus 271, 350-359. 
\item[--] Joy, K.~H., Zolensky, M.~E., Nagashima, K., Huss, G.~R., Ross, D.~K., McKay, D.~S., Kring, D.~A.\ 2012.\ Direct Detection of Projectile Relics from the End of the Lunar Basin-Forming Epoch.\ Science 336, 1426. 
\item[--] Kamata, S., and 13 colleagues 2015.\ The relative timing of Lunar Magma Ocean solidification and the Late Heavy Bombardment inferred from highly degraded impact basin structures.\ Icarus 250, 492-503. 
\item[--] Kring, D.~A., Cohen, B.~A.\ 2002.\ Cataclysmic bombardment throughout the inner solar system 3.9-4.0 Ga.\ Journal of Geophysical Research (Planets) 107, 5009-1. 
\item[--] Kruijer, T.S., Kleine, T., Fischer-Godde, M., Sprung, P., 2015. Lunar tungsten isotopic evidence for the late veneer.  Nature 520, 534–537
\item[--] Laurenz, V., Rubie, D.~C., Frost, D.~J., Vogel, A.~K.\ 2016.\ The importance of sulfur for the behavior of highly-siderophile elements during Earth's differentiation.\ Geochimica et Cosmochimica Acta 194, 123-138. 
\item[--] Levison, H.F., Morbidelli, A., Tsiganis, K., Nesvorny, D., Gomes, R. 2011. Late Orbital Instabilities in the Outer Planets Induced by Interaction with a Self-gravitating Planetesimal Disk. The Astronomical Journal 142, 152.
\item[--] Li, J., Fei, Y. 2014. Experimental Constraints on Core Composition. In Treatise on Geochemistry, 527-557.
\item[--] Li, Y., Aud{\'e}tat, A.\ 2015.\ Effects of temperature, silicate melt composition, and oxygen fugacity on the partitioning of V, Mn, Co, Ni, Cu, Zn, As, Mo, Ag, Sn, Sb, W, Au, Pb, and Bi between sulfide phases and silicate melt.\ Geochimica et Cosmochimica Acta 162, 25-45. 
\item[--] Lock, S.~J., Stewart, S.~T., Petaev, M.~I., Leinhardt, Z.~M., Mace, M., Jacobsen, S.~B., {\'C}uk, M.\ 2016.\ A New Model for Lunar Origin: Equilibration with Earth Beyond the Hot Spin Stability Limit.\ Lunar and Planetary Science Conference 47, 2881. 
\item[--] Lock, S.~J., Stewart, S.~T.\ 2017.\ The structure of terrestrial bodies: Impact heating, corotation limits, and synestias.\ Journal of Geophysical Research (Planets) 122, 950-982. 
\item[--] Mann U., Frost D. J., Rubie D. C., Becker H. and Audétat A. 2012. Partitioning of Ru, Rh, Pd, Re, Ir and Pt between liquid metal and silicate at high pressures and high temperatures  - Implications for the origin of highly siderophile element concentrations in the Earth's  mantle. Geochim. Cosmochim. Acta 84, 593–613.
\item[--] Marchi, S., Mottola, S., Cremonese, G., Massironi, M., Martellato, E. 2009. A New Chronology for the Moon and Mercury. The Astronomical Journal 137, 4936-4948. 
\item[--] Marchi, S., Bottke, W.~F., Kring, D.~A., Morbidelli, A.\ 2012.\ The onset of the lunar cataclysm as recorded in its ancient crater populations.\ Earth and Planetary Science Letters 325, 27-38. 
\item[--] Marchi, S., and 10 colleagues 2013. High-velocity collisions from the lunar cataclysm recorded in asteroidal meteorites. Nature Geoscience 6, 303-307.
\item[--] Marchi, S., Canup, R. M. and Walker, R. J. 2017. Heterogeneous delivery of silicate and metal to the Earth by large planetesimals.  Nature Geoscience, doi:10.1038/s41561-017-0022-3
\item[--] Marinova, M.M., Aharonson, O., Asphaug, E. 2008. Mega-impact formation of the Mars hemispheric dichotomy. Nature 453, 1216-1219. 
\item[--] Maurer, P., Eberhardt, P., Geiss, J., Grogler, N., Stettler, A., Brown, G.~M., Peckett, A., Krahenbuhl, U.\ 1978.\ Pre-Imbrian craters and basins - Ages, compositions and excavation depths of Apollo 16 breccias.\ Geochimica et Cosmochimica Acta 42, 1687-1720. 
\item[--] Mavrogenes, J.~A., O'Neill, H.~S.~C.\ 1999.\ The relative effects of pressure, temperature, and oxugen fugacity on the solubility of sulfide in mafic magmas..\ Geochimica et Cosmochimica Acta 63, 1173-1180. 
\item[--] McCubbin F. M, Vander Kaaden K. E., Tartèse R., Klima R. L., Liu Y., Mortimer J., Barnes J. J., Shearer C. K., Treiman A. H., Lawrence D. L., Elardo S. M., Hurley D. M., Boyce J. W. and Anand 
M. 2015. Magmatic volatiles (H, C, N, F, S, Cl) in the lunar mantle, crust, and regolith: Abundances, 
distributions, processes, and reservoirs. Amer. Mineral. 100, 1668–1707.
\item[--] Meyer, J., Elkins-Tanton, L., Wisdom, J. 2010. Coupled thermal-orbital evolution of the early Moon. Icarus 208, 1-10. 
\item[--] Minarik W. G., Ryerson F. J., and Watson E. B. (1996) Textural entrapment of core-forming melts, Science 272, 530-533.
\item[--] Minton, D.A., Richardson, J.E., Fassett, C.I. 2015. Re-examining the main asteroid belt as the primary source of ancient lunar craters. Icarus 247, 172-190.
\item[--] Morbidelli, A., Levison, H.~F., Tsiganis, K., Gomes, R. 2005. Chaotic capture of Jupiter's Trojan asteroids in the early Solar System. Nature 435, 462-465.  
\item[--] Morbidelli, A., Marchi, S., Bottke, W.F., Kring, D.A. 2012. A sawtooth-like timeline for the first billion years of lunar bombardment. Earth and Planetary Science Letters 355, 144-151.
\item[--] Morbidelli, A., Vokrouhlick{\'y}, D.\ 2003.\ The Yarkovsky-driven origin of near-Earth asteroids.\ Icarus 163, 120-134. 
\item[--] Mungall J. E. and Brenan J. M. (2014) Partitioning of platinum-group elements and Au 
between sulfide liquid and basalt and the origins of mantle-crust fractionation of the 
chalcophile elements. Geochim. Cosmochim. Acta 125, 265–289.
\item[--] Neukum, G., Wilhelms, D.E. 1982. Ancient Lunar Impact Record. Lunar and Planetary Science Conference 13, 590-591. 
\item[--] Neukum, G., Ivanov, B.A. 1994. Crater Size Distributions and Impact Probabilities on Earth from Lunar, Terrestrial-planet, and Asteroid Cratering Data. Hazards Due to Comets and Asteroids 359. 
\item[--] Neukum, G., Ivanov, B.~A., Hartmann, W.~K.\ 2001.\ Cratering Records in the Inner Solar System in Relation to the Lunar Reference System.\ Space Science Reviews 96, 55-86. 
\item[--] Neumann, G.~A., Goossens, S., Head, J.~W., Mazarico, E., Melosh, H.~J., Smith, D.~E., Wieczorek, M.~A., Zuber, M.~T., Lola Science Team, Grail Science Team 2017.\ Lunar Impact Basin Population and Origins Revealed by LOLA and GRAIL.\ New Views of the Moon 2 - Europe 1988, 6037. 
\item[--] Nesvorny, D., Vokrouhlicky, D., Morbidelli, A. 2007. Capture of Irregular Satellites during Planetary Encounters. The Astronomical Journal 133, 1962-1976. 
\item[--] Nesvorny, D., Morbidelli, A. 2012. Statistical Study of the Early Solar System's Instability with Four, Five, and Six Giant Planets. The Astronomical Journal 144, 117. 
\item[--] Nesvorny, D., Vokrouhlicky, D., Morbidelli, A. 2013. Capture of Trojans by Jumping Jupiter. The Astrophysical Journal 768, 45. 
\item[--] Nesvorny, D. 2015a. Evidence for Slow Migration of Neptune from the Inclination Distribution of Kuiper Belt Objects. The Astronomical Journal 150, 73. 
\item[--] Nesvorny, D. 2015b. Jumping Neptune Can Explain the Kuiper Belt Kernel. The Astronomical Journal 150, 68. 
\item[--] Nesvorn{\'y}, D., Roig, F., Bottke, W.~F.\ 2017.\ Modeling the Historical Flux of Planetary Impactors.\ The Astronomical Journal 153, 103. 
\item[--] Nimmo, F., Hart, S.~D., Korycansky, D.~G., Agnor, C.~B.\ 2008.\ Implications of an impact origin for the martian hemispheric dichotomy.\ Nature 453, 1220-1223. 
\item[--] Norman, M.~D., Borg, L.~E., Nyquist, L.~E., Bogard, D.~D.\ 2003.\ Chronology, geochemistry, and petrology of a ferroan noritic anorthosite clast from Descartes breccia 67215: Clues to the age, origin, structure, and impact history of the lunar crust.\ Meteoritics and Planetary Science 38, 645-661. 
\item[--] Norman, M.D. 2009. The Lunar Cataclysm: Reality or “Mythconception”?. Elements, 5, 23-28.    
\item[--] Norman, M.D., Nemchin, A.A. 2014. A 4.2 billion year old impact basin on the Moon: U-Pb dating of zirconolite and apatite in lunar melt rock 67955. Earth and Planetary Science Letters 388, 387-398. 
\item[--] Nyquist, L.~E., Shih, C.-Y.\ 1992.\ The isotopic record of lunar volcanism.\ Geochimica et Cosmochimica Acta 56, 2213-2234. 
\item[--] O'Neill, H.~S.~C.\ 1991.\ The origin of the moon and the early history of the earth - A chemical model. Part 2: The earth.\ Geochimica et Cosmochimica Acta 55, 1159-1172.
\item[--] Pahlevan, K., Stevenson, D.~J.\ 2007.\ Equilibration in the aftermath of the lunar-forming giant impact.\ Earth and Planetary Science Letters 262, 438-449. 
\item[--]  Papanastassiou, D.~A., Wasserburg, G.~J.\ 1971a.\ Rb sbnd Sr ages of igneous rocks from the Apollo 14 mission and the age of the Fra Mauro formation.\ Earth and Planetary Science Letters 12, 36-48. 
\item[--] Papanastassiou, D.~A., Wasserburg, G.~J.\ 1971b.\ Lunar chronology and evolution from Rb sbnd Sr studies of Apollo 11 and 12 samples.\ Earth and Planetary Science Letters 11, 37-62. 
\item[--] Perera, V., Jackson, A.~P., Gabriel, T.~S.~J., Elkins-Tanton, L.~T., Asphaug, E.\ 2017.\ Expedited Cooling of the Lunar Magma Ocean Due to Impacts.\ Lunar and Planetary Science Conference 48, 2524. 
\item[--] Rai, N., van Westrenen, W.\ 2014.\ Lunar core formation: New constraints from metal-silicate partitioning of siderophile elements.\ Earth and Planetary Science Letters 388, 343-352. 
\item[--] Raymond, S.~N., Schlichting, H.~E., Hersant, F., Selsis, F.\ 2013.\ Dynamical and collisional constraints on a stochastic late veneer on the terrestrial planets.\ Icarus 226, 671-681. 
\item[--] Rickman, H., Wi{\'s}niowski, T., Gabryszewski, R., Wajer, P., W{\'o}jcikowski, K., Szutowicz, S., Valsecchi, G.~B., Morbidelli, A.\ 2017.\ Cometary impact rates on the Moon and planets during the late heavy bombardment.\ Astronomy and Astrophysics 598, A67.  
\item[--] Robbins, S.~J.\ 2014.\ New crater calibrations for the lunar crater-age chronology.\ Earth and Planetary Science Letters 403, 188-198. 
\item[--] Rubie, D.C.,  Laurenz, V.,  Jacobson, S.A.,   Morbidelli, A., Palme, H.,  Vogel, A.K.,  Frost, D.J., 2016. Highly siderophile elements were stripped from Earth's mantle by iron sulfide segregation., Science, 353, 1141-1144. 
\item[--] Rushmer, T., Petford, N. 2011. Microsegregation rates of liquid Fe–Ni–S metal in natural silicate-metal systems: a combined experimental and numerical study. Geochem. Geophys. Geosyst.12. 
\item[--] Ryder, G.\ 1990.\ Lunar samples, lunar accretion and the early bombardment of the moon.\ EOS Transactions 71, 313. 
\item[--] Ryder, G. 2002. Mass flux in the ancient Earth-Moon system and benign implications for the origin of life on Earth. Journal of Geophysical Research (Planets) 107, 5022-1.
\item[--] Scheinberg, A., Soderlund, K.~M., Schubert, G.\ 2015.\ Magnetic field generation in the lunar core: The role of inner core growth.\ Icarus 254, 62-71. 
\item[--] Schlichting, H.~E., Warren, P.~H., Yin, Q.-Z.\ 2012.\ The Last Stages of Terrestrial Planet Formation: Dynamical Friction and the Late Veneer.\ The Astrophysical Journal 752, 8. 
\item[--] Singer, K.N., and 22 colleagues 2016. Craters on Pluto and Charon - Surface Ages and Impactor Populations. Lunar and Planetary Science Conference 47, 2310. 
\item[--] Smythe, D.J., Wood, B.J., Kiseeva, E.S. 2017. The S content of silicate melts at sulfide saturation: New experiments and a model incorporating the effects of sulfide composition. American Mineralogist, 102, 795–803.
\item[--] Snape, J.~F., Nemchin, A.~A., Bellucci, J.~J., Whitehouse, M.~J., Tart{\`e}se, R., Barnes, J.~J., Anand, M., Crawford, I.~A., Joy, K.~H.\ 2016.\ Lunar basalt chronology, mantle differentiation and implications for determining the age of the Moon.\ Earth and Planetary Science Letters 451, 149-158. 
\item[--] Solomatov, V.~S.\ 2000.\ Fluid Dynamics of a Terrestrial Magma Ocean.\ Origin of the Earth and Moon 323-338. 
\item[--] Solomatov, V. S. 2015. Magma oceans and primordial mantle di erentiation. Treatise on Geophysics 2nd ed., pp. 81-104.
\item[--] Steenstra, E.S., Rai, N., Knibbe, J.S., Lin, Y.H., van Westrenen, W. 2016. New geochemical models of core formation in the Moon from metal-silicate partitioning of 15 siderophile elements. Earth Planet. Sci. Lett. 441, 1-9.
\item[--] Steenstra, E.S., Lin, Y., Dankers, D., Rai, N., Berndt, J., Matveev, S., Van Westrenen, W., 2017. The lunar core can be a major reservoir for volatile elements S, Se, Te and Sb. Am. Min. and Sci. Rep., 7: 14552.
\item[--] Stegman, D.~R., Jellinek, A.~M., Zatman, S.~A., Baumgardner, J.~R., Richards, M.~A.\ 2003.\ An early lunar core dynamo driven by thermochemical mantle convection.\ Nature 421, 143-146. 
\item[--] Stevenson D.J. 1990. Fluid dynamics of core formation, in the Origin of the Earth, Newsom H E and Jones J H, eds., Oxford Univ. Press, 231-249.
\item[--] St{\"o}ffler, D., Ryder, G.\ 2001.\ Stratigraphy and Isotope Ages of Lunar Geologic Units: Chronological Standard for the Inner Solar System.\ Space Science Reviews 96, 9-54.
\item[--] Strom, R.G., Malhotra, R., Ito, T., Yoshida, F., Kring, D.A. 2005. The Origin of Planetary Impactors in the Inner Solar System. Science 309, 1847-1850. 
\item[--] Suckale, J., Sethian, J.~A., Yu, J.-d., Elkins-Tanton, L.~T.\ 2012.\ Crystals stirred up: 1. Direct numerical simulations of crystal settling in nondilute magmatic suspensions.\ Journal of Geophysical Research (Planets) 117, E08004. 
\item[--] Taylor, S.~R.\ 1975.\ Lunar science - a post-Apollo view. Scientific results and insights from the lunar samples..\ Lunar science - a post-Apollo view.~Scientific results and insights from the lunar samples., by Taylor, S.~R..~ New York, NY (USA): Pergamon Press, 19 + 372 p.
\item[--] Taylor, D.~J., McKeegan, K.~D., Harrison, T.~M., Young, E.~D.\ 2009.\ Early differentiation of the lunar magma ocean . New Lu-Hf isotope results from Apollo 17.\ Geochimica et Cosmochimica Acta Supplement 73, A1317. 
\item[--] Tera, F., Papanastassiou, D.A., Wasserburg, G.J. 1974. Isotopic evidence for a terminal lunar cataclysm. Earth and Planetary Science Letters 22, 1-21.
\item[--] Terasaki, H., Frost, D.~J., Rubie, D.~C., Langenhorst, F.\ 2008.\ Percolative core formation in planetesimals.\ Earth and Planetary Science Letters 273, 132-137.
\item[--] Touboul, M., Puchtel, I.S., Walker, R.J. 2015. Tungsten isotopic evidence for disproportional late accretion to the Earth and Moon. Nature 520, 530-533.  
\item[--] Tsiganis, K., Gomes, R., Morbidelli, A., Levison, H.F. 2005. Origin of the orbital architecture of the giant planets of the Solar System. Nature 435, 459-461. 
\item[--] Turner, G., Cadogan, P.~H., Yonge, C.~J.\ 1973.\ Apollo 17 Age Determinations.\ Nature 242, 513-515. 
\item[--] Walker, R.~J.\ 2009.\ Highly siderophile elements in the Earth, Moon and Mars: Update and implications for planetary accretion and differentiation.\ Chemie der Erde / Geochemistry 69, 101-125. 
\item[--] Walker, R.~J.\ 2014.\ Siderophile element constraints on the origin of the Moon.\ Philosophical Transactions of the Royal Society of London Series A 372, 20130258-20130258. 
\item[--] Vokrouhlicky, D., Pokorny, P., Nesvorny, D. 2012. Opik-type collision probability for high-inclination orbits. Icarus 219, 150-160. 
\item[--] Walsh, K.J., Morbidelli, A., Raymond, S.N., O'Brien, D.P., Mandell, A.M. 2011. A low mass for Mars from Jupiter's early gas-driven migration. Nature 475, 206-209.
\item[--] Walsh, K.~J., Levison, H.~F.\ 2016.\ Terrestrial Planet Formation from an Annulus.\ The Astronomical Journal 152, 68.  
\item[--] Wasserburg, G.~J., Papanastassiou, D.~A.\ 1971.\ Age of an Apollo 15 mare basalt: Lunar crust and mantle evolution.\ Earth and Planetary Science Letters 13, 97-104. 
\item[--] Weber, R.~C., Lin, P.~P., Garnero, E.~J., Williams, Q.~C., Lognonne, P.\ 2011.\ Imaging the Moon's core with seismology.\ AGU Fall Meeting Abstracts . 
\item[--] Werner, S.~C.\ 2014.\ Moon, Mars, Mercury: Basin formation ages and implications for the maximum surface age and the migration of gaseous planets.\ Earth and Planetary Science Letters 400, 54-65. 
\item[--] Wing, B.~A., Farquhar, J.\ 2015.\ Sulfur isotope homogeneity of lunar mare basalts.\ Geochimica et Cosmochimica Acta 170, 266-280. 
\item[--] Wykes, J.L., O'Neil, H.S.C., Mavrogenes, J.A. 2015. The Effect of FeO on the Sulfur Content at Sulfide Saturation (SCSS) and the Selenium Content at Selenide Saturation of Silicate Melts. 56, 1407–1424. 
\item[--] Zhong, S., Parmentier, E.~M., Zuber, M.~T.\ 2000.\ A dynamic origin for the global asymmetry of lunar mare basalts.\ Earth and Planetary Science Letters 177, 131-140.

\end{itemize}

\end{document}